\documentclass[11pt]{JHEP3}

\usepackage{amsfonts}
\usepackage{amsmath}
\usepackage{amsthm}
\usepackage{graphicx}

\usepackage[all]{xy}

\def\braket#1{\mathinner{\langle{#1}\rangle}}
\newcommand{\sbraket}[1]{\lbrack #1\rbrack}

\newcommand{\CP}{\mathbb{CP}}

\newcommand{\ii}{\textrm{i}}
\newcommand{\al}{{\alpha'}}
\newcommand{\dalpha}{{\dot{\alpha}}}
\newcommand{\dbeta}{{\dot{\beta}}}
\newcommand{\dgamma}{{\dot{\gamma}}}
\newcommand{\ddelta}{{\dot{\delta}}}

\newcommand{\ga}{\alpha}
\newcommand{\gb}{\beta}

\newcommand{\eps}{\epsilon}

\newcommand{\gs}{\sigma}

\newcommand{\Nhat}{\hat{N}}

\newcommand{\cP}{\mathcal P}

\newcommand{\cR}{\mathcal R}

\newcommand{\cZ}{\mathcal Z}

\newcommand{\bR}{\mathbb R}

\newcommand{\rank}{\mbox{rank}}

\renewcommand{\Re}{\mbox{Re~}}

\newcommand{\tr}{\mbox{Tr~}}
\newcommand{\be}{\begin{equation}}
\newcommand{\bea}{\begin{eqnarray}}

\newcommand{\ee}{\end{equation}}
\newcommand{\eea}{\end{eqnarray}}

\newcommand{\ret}{\nonumber \\}
\newcommand{\sk}{\vspace{1 em} \noindent}

\theoremstyle{definition}
\newtheorem*{definition}{Definition}

\title{MHV, CSW and BCFW: field theory structures in string theory amplitudes}
\author{Rutger Boels \\ Niels Bohr International Academy, Niels Bohr Institute\\ Blegdamsvej 17, DK-2100 Copenhagen, Denmark}
\author{Kasper Jens Larsen \\ Niels Bohr Institute, Blegdamsvej 17, DK-2100 Copenhagen, Denmark  \\ Department of Physics and Astronomy, Uppsala University, SE-75108 Uppsala, Sweden}
\author{Niels A. Obers \\ Niels Bohr Institute, Blegdamsvej 17, DK-2100 Copenhagen, Denmark }
\author{Marcel Vonk \\ Center for Theoretical Physics, University of the Witwatersrand \\ 1, Jan Smuts Ave,
Johannesburg, South Africa }

\preprint{WITS-CTP-038, UUITP-25/08}

\abstract{
Motivated by recent progress in calculating field theory
amplitudes, we study applications of the basic ideas in these
developments to the calculation of amplitudes in string theory.
We consider in particular both non-Abelian and Abelian open superstring disk amplitudes
in a flat space background, focusing mainly on the four-dimensional case.
The basic field theory ideas under consideration split into three separate categories. In the first, we argue that the calculation of $\al$-corrections to MHV open string disk amplitudes reduces to the determination of certain classes of polynomials. This
line of reasoning is then used to determine the $\al^3$-correction to the MHV
amplitude for all multiplicities. A second line of attack concerns
the existence of an analog of CSW rules derived from the Abelian
Dirac-Born-Infeld action in four dimensions. We show explicitly that
the CSW-like perturbation series of this action is surprisingly
trivial: only helicity conserving amplitudes are non-zero. Last but
not least, we initiate the study of BCFW on-shell recursion relations in
string theory. These should appear very naturally as the UV properties of the string theory are excellent. We show that all open four-point string amplitudes in a flat background at the disk level
obey BCFW recursion relations. Based on the naturalness of the proof
and some explicit results for the five-point gluon amplitude, it is expected that
this pattern persists for all higher point amplitudes and for
the closed string.}

\begin{document}
\section{Introduction}
Recently much technology has become available to calculate
scattering amplitudes in four-dimensional Yang-Mills and gravitational theories. This
was mainly inspired by Witten's twistor string proposal
\cite{Witten:2003nn} and includes at tree level the formulation of
new Feynman-like rules \cite{Cachazo:2004kj} as well as new
recursive relations \cite{Britto:2004ap, Britto:2005fq}. Historically,
the development of new technology in four-dimensional Yang-Mills
theory was often motivated from string theory. However, amplitude
technology has come a long way, and perhaps now is the time to
reverse the reasoning. Specifically, to what extent can we find
structures present in gauge theory such as MHV amplitudes, CSW rules
and BCFW on-shell recursion relations directly for the string?
Interestingly, string theory grew out of `analytic S-matrix'
approaches to field theory, and this is exactly the type of approach
which is dominating the cutting edge of (analytic) amplitude
calculations in the last few years. In this article we take a first
step in applying these new field theory techniques and results to
string theory amplitudes in a three-pronged attack: by studying
analogs of the MHV amplitude, by manipulating the effective
space-time action for Abelian fields and by investigating on-shell
recursion relations.

The MHV or Parke-Taylor \cite{Parke:1986gb} amplitude is the
scattering amplitude in four-dimensional Yang-Mills theory for a
process which has $2$ massless particles with a certain helicity,
say $-$, and an arbitrary number of massless particles with the
opposite helicity. It displays a remarkable simplicity. The question
of $\al$ corrections to this result was first raised in a series of
papers by Stieberger and Taylor \cite{Stieberger:2006bh,
Stieberger:2006te, Stieberger:2007jv}. At the disk level the leading
term in the $\al$ expansion is reproduced simply by Yang-Mills
theory. Even more, at that level the conformal anomalies can safely
be ignored so that one can perform calculations with just a four-dimensional target
space. Here the $5,6$ and $7$-gluon amplitudes were calculated
directly from the string theory. Then, based on soft limits,
Stieberger and Taylor conjectured an all-multiplicity expression for
the first correction of the MHV amplitudes at order $\al^2$. The
result displays a similar simplicity as the Yang-Mills MHV amplitude. However,
it is clear that the worldsheet-dominated methods employed in these
papers in their current form do not really `scale' well with
particle multiplicity, so new input is needed. We propose in this
paper that this input can be motivated from the four-dimensional target
space point of view, with some very basic physical considerations
and string theory computations. In the course of the investigations,
several interesting structures will be uncovered.

Apart from the desire to learn more about string theory, one should
keep in mind that there are several
technical interconnections known between field theory and string
theory which also motivate the investigation reported here. For
example, the integration over Feynman parameters in a generic loop
integral in field theory resembles the integration over vertex
operator insertions in string theory. As another example, the
$\al^2$ correction to the MHV amplitude resembles up to an almost
trivial factor the $1$-loop all plus amplitude
\cite{Stieberger:2006bh}. To connect to more recent work, we show in
this article the (not-so-surprising) result that the string theory
MHV amplitude must be proportional to the field theory  MHV
amplitude\footnote{This was also noticed recently in
\cite{Berkovits:2008ic}.}. However, this is the same reasoning as
used to argue that the $\mathcal{N}=4$ MHV amplitude is proportional
to the tree level MHV amplitude to all loop orders. The
proportionality function there is the subject of much debate as it
can be calculated both at weak (e.g. \cite{Bern:2007ct}) and strong
(e.g \cite{Alday:2007hr}) coupling and has a surprising connection
to Wilson loops, integrability and other interesting structures in
field and string theory. In this context it is intriguing that the
(four-point) proportionality function in the flat background case
considered in this paper obeys for instance a maximal
transcendentality principle.

This article is structured as follows: in section \ref{sec:review}
some general considerations are discussed and some techniques common in
field theory will be shown to apply directly to string theory. These
observations will be put to work in section \ref{sec:alphaprime3},
where we obtain the all-multiplicity MHV amplitude up to order
$\al^3$. In the next section, we switch tracks and study what can be
learned about Abelian amplitudes from studying the Dirac-Born-Infeld
action in four dimensions. Surprisingly, we are able to show
diagrammatically that the DBI action only generates helicity
conserving amplitudes by obtaining very simple CSW-like Feynman rules. In particular, the only MHV amplitude which does not vanish is the one for
$4$ photons. A final section \ref{sec:recursion}
studies the subject of BCFW recursion relations in string theory. We
show that these seem to arise naturally and involve stringy concepts
as `duality', `Regge-behavior' and `resonances' which go back all
the way to the birth of string theory, which ties neatly into the observation in the opening paragraph of this section. The practical use of these
relations is still quite limited at present, but must be seen
as an invitation to join the fun. Conclusions round off the main
presentation, and some technicalities are dealt with in several
appendices.

\section{General considerations for string theory amplitudes}
\label{sec:review} In this section we review various known string
and field theory facts about amplitude calculations and show that
some arguments known in field theory carry over directly to the open
superstring theory setting. The actual amplitudes can be obtained in
basically two different ways: by direct calculation, or by
constructing the effective action through other methods and
calculating the Feynman diagrams. Aside from these calculations
there are constraints on the amplitudes from soft and collinear
limits, and from supersymmetry. As a matter of notation, we will
write gluon amplitudes in `color-ordered' form,
\begin{equation}\label{eq:colorordering}
A_{\textrm{full}}(p_1,p_2,\ldots p_n) = g^{n-2} \sum_{\sigma \in
S_n/\mathbb{Z}_n} A_{\textrm{sub}}(\sigma(1), \ldots, \sigma(n)) \,
\tr(T^{a_{\sigma(1)}} \ldots T^{a_{\sigma(n)}}) \ ,
\end{equation}
where $T^{a_i}$ are the generators of the gauge group (see e.g.
\cite{Dixon:1996wi}). This is of course very natural from the point
of view of the underlying disk diagram. The quantity
$A_{\textrm{sub}}$ is referred to as a subamplitude. It is easy to
see that this amplitude must be cyclic.

\subsection{Direct calculation}
Gluon (or photon) amplitudes in open string theory can of course be
calculated through the usual operator representation, see e.g.
\cite{Green:1987sp}. Vertex operators are inserted along the
boundary of the disk and the position of three of them can be fixed
by global conformal invariance, whereas the rest are integrated
over. Actually, it is these integrations which form in general the
obstacle to calculating the scattering amplitudes in full
generality, and \emph{simple} concrete results through this direct
avenue of attack are limited to the three and for gluon scattering
amplitudes, which read
\begin{equation}\label{eq:threepoints}
A_3(1^-,2^-,3^+) = \frac{\langle 12 \rangle^4}{\langle 12 \rangle
\langle 23 \rangle \langle 31 \rangle} \ ,
\end{equation}
and
\begin{equation}\label{eq:fourpoints}
A_4(1^-,2^-,3^+,4^+) = \frac{\langle 12 \rangle^4}{\langle 12
\rangle \langle 23 \rangle \langle 34 \rangle \langle 41 \rangle}
\frac{\Gamma(1 - \alpha' s) \Gamma(1 - \alpha't)}{\Gamma(1 -
\alpha'(s+t))} \ ,
\end{equation}
in color ordered form. Note that these amplitudes are calculated in
superstring theory; in the bosonic string there will for instance be an extra
contribution ($\mathcal{O}(\al)$) to the three particle amplitude.
The four-point amplitude is of course closely related to the classical
Veneziano amplitude \cite{Veneziano:1968yb}. As an aside: it is
remarkable that the $\al$ expansion of this four-point gluon
superstring amplitude seems to satisfy a `maximal transcendentality'
principle: the dimensionless constant multiplying the $\al^i$
contribution appears to be a product of $\zeta$ functions whose
arguments add to $i$.

Progress for explicit expressions for the higher point amplitudes
has been very limited, although recent work has led to forms for up to
the seven-point amplitude \cite{Medina:2002nk, Oprisa:2005wu, Stieberger:2006bh,
Stieberger:2006te, Stieberger:2007jv} and those which are related to
these by supersymmetry, i.e.\ with up to four gluinos and/or adjoint
scalars as external states. However, further progress along this
line seems hard. The problem seems to be that a high point amplitude
will involve a sum over various integrals over positions of vertex
operators. Both the sum over the integrals as well as the integrals
themselves seem very complicated. However, see appendix A of
\cite{Berkovits:2008ic} for some recent promising progress with
regard to the sum at least.

\subsection{Effective action arguments}
One reason one would like to calculate gluon amplitudes is to
reconstruct the string theory effective action which can then be
taken off-shell to study all kinds of physical effects. However, the
effective action can also be obtained by various other methods. For
Abelian fields, there is for instance the classic result (see
\cite{Tseytlin:1999dj} and references therein) that the effective
action takes the form
\begin{equation}
S[A] = S_{\textrm{DBI}}[A] + S_{\textrm{derivatives}}[A] \ ,
\label{eq:eff-action}
\end{equation}
where the leading-in-derivatives piece is the famous
Dirac-Born-Infeld (DBI) action,
\begin{equation}
S_{\textrm{DBI}} = -1 + \frac{1}{\pi^2 g_s \al^4}\int d^{10}x
\sqrt{-\det\left(\eta_{\mu \nu} + \al \pi F_{\mu \nu} \right)} \ .
\end{equation}
Note that from a perturbative point of view, this action contains an
infinite series of vertices. The sub-leading pieces in
(\ref{eq:eff-action}) contain derivatives acting on the field
strength tensor. In the Abelian case, there is a clean
gauge-independent derivative expansion. This expansion however
breaks down in the non-Abelian case since two derivatives can always
be traded for a field strength tensor, e.g. for an adjoint field $H$,
\begin{equation}
[D_{\mu}, D_{\nu}] H = [F_{\mu\nu},H] \:  \ .
\end{equation}
Therefore in the non-Abelian case one needs to find the complete
effective action in one go, which makes the problem of determining
it much harder. The effective action is, however, known up to order
$\al^3$ \cite{Koerber:2001uu}.

To our best knowledge, deriving scattering amplitudes from string theory effective actions usually proceeds
by laborious Feynman diagram calculations. We will see in section \ref{sec:action}, however, that already in the Abelian case these calculations can be streamlined quite considerably in four dimensions.

\subsection{Analytic constraints}\label{sec:analytic-constraints}
Since explicit computation of string theory gluon scattering
amplitudes seems to be hard by either method mentioned above, let us
focus on the analytic behavior of the amplitudes as momentum
invariants vanish. For color-ordered amplitudes as written in
\eqref{eq:colorordering}, the quantity $A_{\textrm{sub}}$ has a
certain specific ordering of the gluons. As a consequence, the
sub-amplitude can only have poles if a momentum invariant
constructed out of consecutive gluon momenta vanishes. That is, the
amplitudes can only have poles of the form
\begin{equation}
\textrm{poles} \sim \frac{1}{\left(p_i + p_{i+1} + \ldots + p_j
\right)^2 + \frac{k}{\al}} \ ,  \label{eq:poles}
\end{equation}
for some mass $m_l^2 = \frac{k}{\al}$ with $k$ a non-negative
integer. In particular, it can be zero.

\begin{figure}[t]
  \begin{center}
  \includegraphics[scale=0.6]{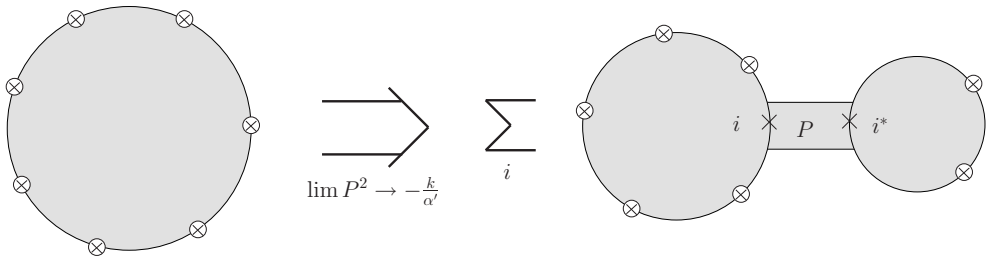}
  \caption{Conformal symmetry elucidates a certain kinematical limit with $k$ a non-negative integer. The sum runs over all string states at this particular mass level. }
  \label{fig:factorizationprops}
  \end{center}
\end{figure}

Actually, the form of the poles is easy to see directly from the
disk diagram. As indicated in figure \ref{fig:factorizationprops},
we can always, by conformal transformations, map the disk into a
structure which has a very long strip stretching between two disks.
As the strip between the disks becomes very long, the states which
can travel between the disks start to become closer and closer to
the mass-shell. This will lead through the string theory propagator
to poles of the form indicated above in (\ref{eq:poles}). In fact,
for a general amplitude there can be infinitely many different
massive states which can travel through the strip corresponding to
the full tower of states of the string. It is easy to see that the
residue at a certain pole where $\left(p_1 + p_2 + \ldots + p_j
\right)^2 \rightarrow - \frac{k}{\al}$ will be exactly two disk
amplitudes,
\begin{equation}
A_n(1,2,3 \ldots, n) \rightarrow \sum_i \frac{A_{j+1}(1,2,\ldots, j,
i) A_{n-j+1}(i, j+1, \ldots, n)}{\left(p_1 + p_2 + \ldots + p_j
\right)^2 + \frac{k}{\al}} \label{eq:residue} \ ,
\end{equation}
where one has to sum over all the different intermediate states of
the string at a certain mass level. Note that this sum splits into a
sum over different particles and their quantum numbers (such as
angular momentum). A discussion of this can be found in e.g.
\cite{Green:1987sp}.

As a special case, one can consider the massless poles of a gluon
amplitude. These are interesting since the residues at these poles
are again amplitudes which only contain gluons. When a three-gluon
disk factors out, so in the kinematic limit
\begin{equation}\label{eq:colllimit}
(p_j + p_{j+1})^2 = 2 p_j \cdot p_{j+1} \rightarrow 0 \:  \ ,
\end{equation}
for some index $j$ equation \eqref{eq:residue} simplifies
considerably. In this particular case, the residue of this kinematic
pole consists of a three gluon superstring disk diagram
(\ref{eq:threepoints}) times an $n-1$ gluon disk diagram. Since the
three gluon diagram is actually independent of $\al$, it is seen
that the general structure of the residue is precisely the same as
in tree level Yang-Mills theory. In particular, this implies that
the superstring theory has the \emph{exact same} collinear
singularities as the field theory\footnote{Note that in the case of
the bosonic string, there is a $\mathcal{O}(\al)$ correction to
equation (\ref{eq:threepoints}) which spoils this argument.}. Note
that the technical difficulty that the three-point Yang-Mills
amplitudes vanish in Minkowski signature can easily be avoided by
generalizing to complex momenta. The singularities can conveniently
by described by the so-called splitting functions (see e.g.
\cite{Dixon:1996wi}). Let $S(i^\pm,j^\pm,(j+1)^\pm)$ denote the
splitting function with intermediate state $i$, and external states
$j$ and $j+1$ and the indicated helicity. In the collinear limit
\eqref{eq:colllimit},
\begin{equation}
p_j^{\mu} = z Q^{\mu} \quad , \quad p_{j+1}^{\mu} = (1-z) Q^{\mu}
\quad , \quad Q^2=0 \ ,
\end{equation}
the splitting functions are easily derived from, for instance, the
MHV amplitude and read
\begin{equation}\label{eq:splitfunc}
\begin{array}{ccc}
S(i^+ j^+ (j+1)^+) = 0 & \quad  ,\quad \quad & S(i^- j^+ (j+1)^+) = \frac{1}{\sqrt{(1-z) z}} \frac{1}{\braket{j, j+1}}  \\
S(i^+ j^+ (j+1)^-) = \frac{(1-z)^2}{\sqrt{(1-z) z}} \frac{1}{\braket{j, j+1}} & ,\quad & S(i^- j^+ (j+1)^-) =  \frac{(1-z)^2}{\sqrt{(1-z) z}} \frac{1}{\sbraket{j, j+1}} \\
S(i^+ j^- (j+1)^-) = \frac{1}{\sqrt{(1-z) z}} \frac{1}{\sbraket{j,
j+1}} & ,\quad & S(i^- j^- (j+1)^-) = 0 \ .
\end{array}
\end{equation}

An even simpler singularity arises in the case where one of the
momenta vanishes, say $p_j \rightarrow 0$. This so-called `soft'
singularity gives rise to a pole which is closely related to the
above one, since a soft singularity is of course a special case of a
collinear singularity. This is discussed in e.g chapter $6$ of \cite{Mangano:1990by}.

\subsection{Effective supersymmetry}\label{sec:eff-susy}
As noted in \cite{Dixon:1996wi}, the form of the MHV amplitude in
four-dimensional field theory can be understood at tree level
through an effective supersymmetry. In particular, the
supersymmetric Ward identities determine in these four dimensions
\cite{Grisaru:1977px}
\begin{equation}\label{eq:vanishingamps}
A(++\ldots+) =0 \: , \quad \quad A(++\ldots+-) =0 \:  ,
\end{equation}
with an exception only for the three particle $A(++-)$ and $A(--+)$
amplitude which vanish generically only for all momenta in
$\bR^{(1,3)}$. The input in this argument is nothing more than simple
representation theory of \emph{on-shell} $\mathcal{N}=1$ space-time
supersymmetry and the absence of helicity-violating fermion
amplitudes \footnote{The absence of helicity-violating fermion
amplitudes in open superstring theory follows from charge
conjugation invariance. To see this note that the vertex operators
of space-time fermions involve spin fields $\Theta_\alpha$
satisfying the OPE $\Theta_\alpha (z) \Theta_\beta (w) \propto
C_{\alpha \beta}$ where $C$ is the charge conjugation matrix, which
is non-zero only for spinors of opposite chirality.}. Just as the
derivation of these Ward identities is independent of the loop
counting parameter $\hbar$ in loop computations in ordinary
supersymmetric Yang-Mills theory, it is actually independent of
$\al$ (and $g_s$ for that matter). Hence \eqref{eq:vanishingamps}
hold to arbitrary order in $\al$, again in four dimensions. Note that these are the only amplitudes in open superstring theory which are known to all orders. This is the target space counterpart of the worldsheet based argument in
\cite{Stieberger:2007jv}.

Equation \eqref{eq:vanishingamps} implies that the next helicity amplitude in line, the MHV
amplitude, cannot have multi-particle massless kinematic poles: on such poles
the amplitude would factorize into two pieces, one of which is bound
to vanish. This is because the propagator always respects helicity
and with our `ingoing' helicity convention, it is of the form
`$+-$'. It is therefore clear that it has only two-particle
collinear poles. These have a universal form,
\begin{equation}
A(+\ldots -_i \ldots -_j \ldots +) \sim
\frac{D^{(N)}_{ij}}{\braket{1 2} \braket{2 3} \cdots \braket{N 1}} \ ,
\label{eq:universalform}
\end{equation}
which can easily be checked to encode all the collinear poles in the splitting functions in equation \eqref{eq:splitfunc}. Effective $\mathcal{N}=4$ supersymmetry also requires that the numerator of this amplitude
reads
\begin{equation}
D^{(N)}_{ij} = \braket{i j}^4 Q^{(N)} \ ,
\end{equation}
where $Q^{(N)}$ is a function of the momenta. More precisely,
supersymmetry requires that
\begin{equation}
\frac{A(+\ldots -_i \ldots -_j \ldots +)}{\braket{i j}^4} =
\frac{A(+\ldots -_i \ldots -_{j+1} \ldots +)}{\braket{i (j+1)}^4} \ ,
\end{equation}
so that this particular combination does not depend on $i$ or $j$.
Dimensional analysis now fixes $Q^{(N)}$ to be (mass-)dimensionless
and have spinor weight zero. Note incidentally that this is the same
argument as is used in loop calculations in $\mathcal{N}=4$
Yang-Mills theory to argue that the MHV amplitude is proportional to
the tree amplitude to all loop orders. The analog of the
proportionality function $Q^{(N)}$ in that case is the subject of
much ongoing debate because it can be calculated at both strong and weak coupling.

In the case under study here, since all the massless kinematic poles
of the amplitude have been expressed in \eqref{eq:universalform} and
we are considering the effective theory of massless gluons only,
$Q^{(N)}$ has to be a polynomial in the momenta. There can be no further kinematic poles as there are no more propagating particles. At the classical
level in pure Yang-Mills theory there is no natural parameter
with the dimension of mass. This fixes $Q^{(N)}$ to be a particle number independent constant in
this case, which can then be determined by considering the $3$-point amplitudes. On the other hand, in string theory there is a dimensionful constant $\alpha'$ of
dimension $-2$. In general the function $Q^{(N)}$ will have numerous
poles for the exchange of \emph{massive} particles. However, if we
consider an $\al$ expansion, these will have been integrated out, so
in the effective theory there are only massless propagating
particles. Therefore when considering such an expansion, $Q^{(N)}$
is again a polynomial in the external momenta.

Explicitly, $Q^{(N)}$ may be expanded in $\al$ as
\begin{equation}\label{eq:Qexpansion}
 Q^{(N)} = \sum_{n \geq 0} (\ga')^n Q^{(N)}_n \ ,
\end{equation}
where $Q^{(N)}_n$ are homogeneous polynomials in the momenta of the
external particles of mass dimension $2n$. The coefficients
$Q^{(N)}_n$ in the color ordered amplitude now inherit certain
properties from the parent amplitude. We will see in section
\ref{sec:alphaprime3} how these properties and some known results
for low external particle numbers fix the form of the $\al$
corrections to the MHV amplitude for any number of external
particles up to order $\al^3$.

\subsection*{Twistor space interpretation}
The above form of the string theory MHV amplitude can be summarized
by the following statement: in the low energy limit the ordinary
superstring theory MHV amplitudes at the disk level localize on a
holomorphic line with derivative of delta function support. This
follows easily from Witten's original insight \cite{Witten:2003nn},
see also \cite{Cachazo:2005ga}. In split $(2,2)$ signature this can
be seen by considering the amplitude to be a functional of all the
spinor momenta,
\begin{equation}
A\left(p^{i}_\dalpha, p^{i}_{\alpha}, \ldots \right) \sim
\delta\Big(\sum_{i} p^i \Big) \frac{\braket{rs}^4 \left(
Q(p^{i}_\dalpha, p^{i}_{\alpha}) \right)}{\braket{1 2} \braket{2 3}
\ldots \braket{n 1}} \: ,
\end{equation}
temporarily resurrecting an otherwise suppressed momentum conserving
delta function. In split signature a lift to twistor space boils
down to performing a Fourier transform with respect to all the
undotted spinors,
\begin{equation}\label{eq:transftotwistorspacesplit}
\tilde{A} \left(p^{i}_\dalpha,  \tilde{p}^{i}_{\alpha}, \ldots
\right) = \int \left(\prod_j dp^j_\alpha \exp^{i
\left(\sbraket{\tilde{p}^{i} p^{j}} \right)}  \right)
A\left(p^{i}_\dalpha, p^{j}_{\alpha}, \ldots \right) \ .
\end{equation}
The momentum conserving delta function can be represented as an
integral through
\begin{equation}
\delta\Big(\sum_{i} p_i \Big) = \int d^4x \exp^{i\left( x^{\alpha
\dalpha} \sum_{i} p^i_{\alpha} p^i_{\dalpha} \right)} \ .
\end{equation}
The only dependence on the anti-holomorphic spinors $p^i_{\alpha}$
in our general form of the amplitude are contained within the
polynomial in the numerator. By the Fourier transform in
(\ref{eq:transftotwistorspacesplit}), these occurrences can always
be replaced by derivatives by the usual rule,
\begin{equation}
p^i_{\alpha} \rightarrow - i \frac{\partial}{\partial
\tilde{p}^i_{\alpha}} \ .
\end{equation}
The remaining integral over $p^i_{\alpha}$ can then easily be
recognized as (a series of) delta function(s). The twistor space
function $\tilde{A}$ will therefore have the form
\begin{equation}
\tilde{A} = Q\left(p^{i}_\dalpha, - i \frac{\partial}{\partial
\tilde{p}^i_{\alpha}}\right) \int d^4x \delta(\tilde{p}^i_{\alpha} -
x_{\alpha \dalpha} p^{\dalpha}) A_{\textrm{MHV}} \ .
\end{equation}
The argument of the delta functions is of course the famous
incidence relation in twistor theory. This relation describes the
holomorphic embedding of a $\CP^1$ (line) inside $\CP^3$,
parametrized by the space-time coordinate $x$. Note that the $\al$
corrections imply that the amplitude has derivative of
delta-function support on this line, with twice the number of
derivatives as powers of $\al$.

Localization on a line in twistor space implies that in some sense
the string theory MHV amplitudes are local in space-time. In field
theory this observation was originally made by Nair
\cite{Nair:1988bq}. Since these amplitudes are local one may wonder
whether there is an action which has these amplitudes directly as
vertices. In the field theory this can be derived from the usual
Yang-Mills theory \cite{Boels:2006ir} by lifting to twistor space
(see \cite{Boels:2008ef} for a review), and in the Abelian case this
will be discussed briefly in section \ref{sec:action}.

\section{Determining $\al$ corrections to the MHV amplitude}
\label{sec:alphaprime3} In the previous section it was argued that
the form of the MHV superstring amplitude is determined up to a
polynomial from kinematic and supersymmetric considerations. These
reduce the calculation of this class of amplitudes to classifying
certain polynomials. In the present section we determine these
polynomials for the $\al^3$ correction to the MHV amplitude for any
number of external gluons.

\subsection{Symmetry constraints on the amplitude}
In section \ref{sec:eff-susy} it was shown that the kinematic and
supersymmetric constraints determine that the general
$\ga'$-corrected MHV amplitude with $N$ external gluons must have
the form \be\label{eq:generalMHVansatz}
 A_N ( -+ \cdots-_j + ) =  A_{YM}^{(N)} \left(\sum_{n \geq 0} (\ga')^n Q^{(N)}_n\right).
\ee Here, $N$ is the number of external states in the amplitude.
Every $Q^{(N)}_n$ is a Lorentz invariant polynomial in the momenta
$k^\mu_i$ of the external particles. To make the prefactor
dimensionless, $Q^{(N)}_n$ must be homogeneous and have degree
$d=2n$ in the momenta $k^\mu_i$.

The purpose of this section is to determine how much we can learn
about the polynomials $Q^{(N)}_n$ using their symmetry properties
and the results that are already known in the literature. To this
end, let us list some important properties of the $Q^{(N)}$.
\begin{itemize}
 \item
  Since we study color-ordered amplitudes (see e.g.\ \cite{Mangano:1990by, Dixon:1996wi}), the $Q_n^{(N)}$ are cyclically symmetric in the external momenta $k^\mu_i$.
 \item
  From the fact that the three-point functions for massless poles are $\ga'$-independent (see equation \eqref{eq:threepoints}) it follows that,
  just like in the Yang-Mills case, $Q_n^{(N)}$ must reduce to $Q_n^{(N-1)}$ under soft limits:
  \be
   Q_n^{(N)} \bigg|_{p^\mu_N = 0} = Q_n^{(N-1)} \ .
  \ee
 \item
  For the same reason, the behavior under collinear limits generalizes from the Yang-Mills case to the stringy case. That is $Q^{(N)}$ must reduce to $Q^{(N-1)}$ under collinear limits:
  \be
   Q_n^{(N)} \bigg|_{p^\mu_N = \ga p^\mu, p^\mu_{N-1} =
   (1-\ga) p^\mu} = Q_n^{(N-1)} \bigg|_{p^\mu_{N-1} = p^\mu} \ .
  \ee
\end{itemize}
The soft limit is nothing but the special case $\ga=0$ of the
collinear limit, but it is this special case that we are primarily
interested in. Note in particular that after we apply a soft limit
to a $Q^{(N)}_n$, the resulting answer must again be cyclically
symmetric in the remaining $p_i^\mu$. This leads us to the following
definition:

\begin{definition}
A {\bf cyclically reducible polynomial} of degree $d$ in $N$
variables is a Lorentz invariant polynomial which is cyclically
symmetric in the $N$ variables $p_i^\mu$, and which, after setting a
subset \be \Big\{ p^\mu_i | i \in S \subset \{1, \ldots, N \} \Big\} \ ,
\ee of the variables to zero, is cyclically symmetric in the
remaining variables.
\end{definition}

Here, `cyclically symmetric' does not necessarily mean that the form
of the polynomial is the same after a cyclic permutation of the
indices. In general, there will be relations between the momenta,
such as the momentum conservation condition $\sum p_i^\mu = 0$. By
cyclic symmetry, we mean that after a cyclic permutation of the
indices, the polynomial comes back to itself up to an expression
which vanishes due to these relations.

Our goal will be to find all possible cyclically reducible
polynomials of a given degree $d$ in a given number of external
particles $N$. Since such polynomials form a vector space, it will
be enough to find a basis for this space. As we will see, when $d$
is not too large, these vector spaces have a relatively low
dimension which moreover quickly becomes constant when we increase
$N$. Our strategy is then to express the known $Q^{(N)}_n$ from the
literature in terms of this basis. It will turn out that with a
smart choice of basis, the coefficients of $Q^{(N)}_n$ also become
$N$-independent, meaning that we can find a general expression for
the $(\ga')^n$-term in (\ref{eq:Qexpansion}) which is valid for all
$N$. We will carry out this procedure in detail for $n=2$ (for which
an answer is known in the literature) and $n=3$ (for which our
answer is new).

As a final remark, notice that we do not use collinear limits in the
construction of our answer. It would be interesting to reduce the
basis of `allowed' polynomials even further by requiring good
behavior under collinear limits. For our current purposes, however,
this is unnecessary, since our bases of polynomials are small enough
to handle as they are. In fact, we will now be able to use collinear
limits as a nontrivial test on our results: after constructing our
final answer, we will check explicitly that it behaves as expected
under collinear limits.

\subsection{A toy model: scalar variables}
\label{sec:scalarvariables} Before we turn to our actual problem,
let us develop some intuition by considering a simpler problem. To
this end, we replace the momenta $p_i^\mu$ by scalar variables,
which we will denote by $x_i$. The problem of finding all cyclically
reducible polynomials can now be solved exactly, without the use of
a computer.

\subsubsection*{Basis polynomials}

To begin with, let us also assume that there are no relations
between the $x_i$ -- we will drop this requirement in a moment. The
question as to which polynomials of degree $d$ are cyclically
reducible now has a simple answer:

\sk {\bf Claim.} A basis for the vector space of cyclically
reducible polynomials of degree $d$ in $N$ variables consists of the
polynomials of the form \be
 \label{eq:scalarbasis}
 (a_1 a_2 \ldots a_m) \equiv \sum_{ (i_1, \ldots i_m) } x_{i_1}^{a_1} \cdots x_{i_m}^{a_m} \ ,
\ee where $m \leq N$, $a_i \geq 1$ are positive integers with $\sum
a_i = d$, and the sum is over all cyclically ordered sequences
$(i_1, \ldots i_m)$ of length $m$. Here, $(a_1 \ldots a_m)$ that are
the same up to a cyclic permutation are considered to be the same
basis vector.

\sk By a `cyclically ordered sequence' $(i_1, \ldots i_m)$ we mean
that after identifying $N+1$ with $1$, the sequence has `winding
number' +1. In other words, after an appropriate cyclic permutation,
the sequence will be in a strictly increasing order. Thus, for
example, $(1,2,3,4,5,6)$, $(1,2,3,5,8,9)$ and $(3,6,7,9,1,2)$ are
all cyclically ordered sequences for $d=6$ and $N=9$, but
$(1,3,5,6,8,2)$ and $(6,5,4,3,2,1)$ are not. In appendix
\ref{app:proofbasis}, we prove the claim that (\ref{eq:scalarbasis})
forms a basis of cyclically reducible polynomials.

As an example, when $d=4$ and $N=4$, we have the five basis
polynomials \be
 (4), (31), (22), (211), (1111) \ ,
\ee where for instance \bea
 (211) & = & \phantom{+} x_1^2 x_2 x_3 + x_1^2 x_2 x_4 + x_1^2 x_3 x_4 \ret
&& + x_2^2 x_3 x_4 + x_2^2 x_3 x_1 + x_2^2 x_4 x_1 \ret && + x_3^2
x_4 x_1 + x_3^2 x_4 x_2 + x_3^2 x_1 x_2 \ret && + x_4^2 x_1 x_2 +
x_4^2 x_1 x_3 + x_4^2 x_2 x_3 \ . \eea
Note that if we would have
chosen $d=4$ but $N=3$, the polynomial $(1111)$ would vanish since
there are simply not enough different variables to write down a
term. This is the reason for the constraint $m \leq N$ in our
description above. The fact that we need at least $d$ particles to
write down the full basis of degree $d$ polynomials will become
crucial later on. In general, unless otherwise stated, we will
assume that there are enough particles in the amplitude so that the
constraint $m \leq N$ does not come into play.

\subsubsection*{Number of basis polynomials}

To count the number of basis polynomials of a given degree $d$
(assuming as we mentioned above that $N$ is large enough), we need
to count the number of sequences of positive integers that add up to
$d$, up to cyclic permutations. In mathematics, sequences of
integers up to cyclic permutations are called {\em necklaces}. The
numbers $N_d$ of necklaces of arbitrary length with positive integer
entries that add up to a given $d$ can be found in Sloane's on-line
encyclopedia of integer sequences \cite{Sloane:seq} as A008965; it
starts off as \be
 1, 2, 3, 5, 7, 13, 19, 35, 59, 107, 187, 351, 631, 1181, \ldots
 \ .
\ee In fact, with some smart combinatorics (see appendix
\ref{app:necklaces}) an exact expression for this sequence can be
found: it is \be
 N_d = -1 + \frac{1}{d} \sum_{d_i | d} \phi(d_i) 2^{d/d_i} \ ,
\ee where the sum is over all divisors $d_i$ of $d$. Here,
$\phi(d_i)$ is Euler's totient function, counting the number of
integers $k \leq d_i$ that are relatively prime to $d_i$: \be
 \phi(d_i) = d_i \prod_{p | d_i} (1-p^{-1}) \ ,
\ee where the product is over all different primes $p$ dividing
$d_i$. 
The leading term in this expression tells us that \be
 N_d \sim \frac{2^d}{d} \ ,
\ee is a good estimate for the number of basis polynomials at large
$d$.

\subsubsection*{Momentum conservation}

Now, let us consider the somewhat more realistic problem where there
are relations between the generating monomials. In particular, let
us study the momentum conservation constraint\footnote{In the scalar
case, we do not yet include the analog of masslessness constraint,
$x_i^2 = 0$, since it would oversimplify the problem. Another type
of constraint that we will meet in the vector case -- the Schouten
identity -- also has no analog in the scalar model.} \be
 \sum x_i = 0 \ .
\ee This constraint means that any two polynomials $P$, $P'$ of
degree $d$ represent the same physical quantity if \be
 P = P' + \Big(\sum x_i \Big) B \ ,
\ee for some polynomial $B$ of degree $d-1$. This defines an
equivalence relation on our space of polynomials; a polynomial $P$
is now cyclically reducible if after a cyclic permutation of its
variables, it is mapped to a polynomial $P'$ in the same equivalence
class $[P]$, and if this property is preserved under soft limits.

Clearly, under this new definition, the polynomials
(\ref{eq:scalarbasis}) are still cyclically reducible. Moreover,
following the same reasoning as in appendix \ref{app:proofbasis}, it
can be shown that they cannot split into a sum $P=P_1 + P_2$ of
smaller cyclically reducible polynomials. Since every possible
monomial appears in one of the basis polynomials, this means that
the basis we constructed before still spans the space of all
possible cyclically reducible polynomials. However, it is no longer
a {\em basis}, since there will be relations between the generating
polynomials.

To see when this is the case, let us assume that some linear
combination of the basis polynomials $P_j^{(d)}$ of degree $d$ is
zero up to relations: \be
 \sum_j c_j P_j^{(d)} = \Bigg( \sum_i x_i \Bigg) B^{(d-1)} \ .
\ee From the permutation symmetry properties of this expression, it
is easily seen that this means that $B^{(d-1)}$ must itself be a
linear combination of polynomials of the form (\ref{eq:scalarbasis})
of degree $d-1$. 
It follows that the set of all relations is spanned by expressions of the form
\be
 \Bigg( \sum_i x_i \Bigg) P^{(d-1)}_j \ ,
\ee where $P^{(d-1)}_j$ form a basis of cyclically reducible
polynomials of degree $d-1$. Since the $P^{(d-1)}_j$ are linearly
independent, so are the relations that they generate.

For clarity, let us again look at an example. For $d=5$, consider
the cyclically reducible polynomial \be
 (221) = \sum_{i<j<k} x_i^2 x_j^2 x_k \ ,
\ee where the sum is over cyclic orderings. Multiplying this with
the momentum conservation polynomial \be
 (1) = \sum_i x_i \ ,
\ee we find for $d=6$ the relation \bea
 0 \sim (1) (221) & = & \bigg( \sum_i x_i \bigg) \bigg( \sum_{j<k<l} x_j^2 x_k^2 x_l \bigg) \ret
 & = & \sum_{i<j<k} x_i^3 x_j^2 x_k + \sum_{i<j<k} x_i^2 x_j^3 x_k + \sum_{i<j<k} x_i^2 x_j^2 x_k^2 \ret
 && + \sum_{i<j<k<l} x_i^2 x_j x_k^2 x_l + 2 \sum_{i<j<k<l} x_i^2 x_j^2 x_k x_l \ret
 & = & (321) + (231) + (222) + (2121) + 2 \cdot (2211) \ .
\eea We see that the algebra is very simple: one just adds the 1 to
any of the entries in $(221)$ or inserts a new 1 between two
indices. This procedure should be carried out for the entire set of
generating cyclically reducible polynomials for $d=5$ to get all
relations between the generating cyclically reducible polynomials
for $d=6$.

We have thus found that a true basis of cyclically reducible
polynomials has size \be
 \Nhat_{d} = N_d - N_{d-1} \sim \frac{2^{d-1}}{d} \ ,
\ee a sequence which begins with \be
 0, 1, 1, 2, 2, 6, 6, 16, 24, 48, 80, 164, 280, 550, \ldots  \ ,
\ee where we defined $N_{-1} = 1$ to obtain the correct answer
$\Nhat_{1}=0$ for $d=1$.

\subsection{Vector variables}
\label{sec:vectorvariables} As we have seen, constructing a basis of
cyclically reducible polynomials and calculating its cardinality is
a problem that can be solved explicitly in the scalar variable case.
Unfortunately, such an analytic solution is much harder when the
variables have a vector index. Here, of course, we require that the
resulting polynomial is Lorentz invariant. The naive approach would
be to simply construct the same generating polynomials as in the
previous section, attach indices to the variables, and contract
these indices with metric and epsilon tensors. There are, however,
several reasons why this approach is not straightforward; let us
mention two important ones.

\subsubsection*{Failure of the naive basis construction}

The first reason why the naive approach fails is that not all basis
vectors can be constructed in this way. To see this consider, for
$N=5$ particles, the term \be
 \eps_{1234} \equiv \eps_{\mu \nu \rho \gs} p_1^\mu p_2^\nu p_3^\rho p_4^\gs \ .
\ee Note that, in principle, we can also define this term for $N=4$
particles, but in that case it would be zero due to momentum
conservation. If we would naively construct our basis of cyclically
reducible polynomials in the same way as in the previous section, we
would also include a term $\eps_{1235}$ in the polynomial that
contains $\eps_{1234}$. But from momentum conservation for five
particles, we see that \be
 \eps_{1234} + \eps_{1235} = 0 \ .
\ee In this way, all terms in the naively constructed polynomial
cancel pairwise, and the entire polynomial vanishes. Nevertheless,
there {\em is} a cyclically reducible polynomial consisting of terms
of this type; it can be written in a manifestly cyclically symmetric
way as \be
 \eps_{1234} + \eps_{2345} + \eps_{3451} + \eps_{4512} + \eps_{5123} \ ,
\ee or, since all of these terms are equal due to momentum
conservation and antisymmetry, simply as $\eps_{1234}$. Thus, we see
that our naive construction fails. The (vanishing) polynomial
constructed from $\eps_{1234}$ using our naive method is of course
still cyclically reducible, but it can now be written as a sum of
smaller (nonvanishing but mutually cancelling) cyclically reducible
polynomials. The reason our proof in appendix \ref{app:proofbasis}
fails in this case is that it relies on the fact that monomials do
not vanish when we take a soft limit in an index that does not
appear explicitly.

\subsubsection*{Relations between relations}

Another technical difficulty in the case of vector variables is the
fact that the relations one constructs are no longer independent. To
see this, let us introduce the notation \be
 s_{ij} = g_{\mu \nu} p_i^\mu p_j^\nu \ .
\ee These variables satisfy $s_{ij} = s_{ji}$ and, by masslessness,
$s_{ii}=0$ (no summation), so for $N$ particles there are $N(N-1)/2$
of them. Momentum conservation now leads to $N$ relations between
the $s$-variables of the form \be
 \sum_i s_{ij} = 0, \qquad 1 \leq j \leq N \ .
\ee If we would be considering $d=4$, this would then give $N^2
(N-1) / 2$ relations \be
 r_{jkl} \equiv \sum_i s_{ij} s_{kl} = 0, \qquad 1 \leq j \leq N, \qquad 1 \leq k<l \leq N \ .
\ee Here, the restriction $k<l$ is only not to overcount, but we
define $r_{jkl}$ for $k>l$ similarly. If we now sum over all $k$ at
a given $j$ and $l$ we obtain \be
 \sum_{k} r_{jkl} = \sum_{i,k} s_{ij} s_{kl} = \sum_i r_{lij} \ .
\ee Thus, for each choice of $(j,l)$, we have found a linear
relation between a subset of the relations. We need to correct for
such `relations between relations' when counting the number of
linearly independent basis polynomials.

\subsubsection*{Generic counting of polynomials}
Even though we have not found a simple closed formula for the number
of cyclically reducible polynomials of given degree in the vector
variable case, let us mention that we can at least find a simple
formula for the number of contractions of $2n$ four-dimensional
Lorentz indices modulo Schouten identities. As explained in appendix
\ref{app:countinv} this is given by the $n$'th Catalan's number
squared. Note that this number grows exponentially for large $n$ as
\begin{equation}
C_n =(c_n)^2 \sim \frac{2^{4n}}{\pi n^3} \quad \textrm{for}
\hspace{2mm} n \rightarrow \infty \ .
\end{equation}
Although the explicit forms of the contractions themselves can also
straightforwardly be worked out, they will not be needed in this
paper.

\subsubsection*{Computer approach}
Taking all of the subtleties mentioned above into account in the
analytic approach is a very tedious excercise, though probably not
impossible. We will leave this for future work though, and for the
moment let a computer do the hard work of counting Lorentz invariant
cyclically reducible basis polynomials for vector variables. From
the scalar variable example, we expect that the number of basis
polynomials we will find is constant above a given number of
particles, which will be of the order $N \sim d$. Moreover, we
expect the basis to be expressible in a form which is independent of
the number of particles, as in our $(a_1 \ldots a_m)$ notation
above. We can therefore try to solve the problem for given $d$ and
small $N$ using a computer, and then conjecture that once the answer
stabilizes in terms of $N$, we have found our general answer. It is
this approach that we now turn to.

\subsection{Methodology}
We have implemented our computer construction of a basis of
cyclically reducible polynomials of given degree using Mathematica.
For definiteness, let us describe the case of degree six; the
generalization to other degrees is straightforward.

For $d=6$, there are two ways to contract the Lorentz indices of the
six momentum vectors appearing in a given term: either by three
metric tensors or by one $\eps$- and one metric tensor. In the
notation used above, the first contraction will lead to terms of the
schematic form $sss$; the second contraction to terms of the
schematic form $\eps s$. Note that there will be no relations mixing
terms of these different forms, so we can construct a basis of
cyclically reducible polynomials for each form separately, and then
combine the answers in the end. Here, for illustrative purposes, we
will discuss the $\eps s$ polynomials -- the case of $sss$
polynomials is treated in a completely analogous manner.

Suppose now that we have $N$ external particles. We work by
induction, so we assume that we already know a basis of cyclically
reducible polynomials\footnote{Actually, as we will see below, it is
technically more convenient to store a basis of polynomials {\em
perpendicular} to the space of cyclically reducible ones.} for $N-1$
particles. In this particular example we can start the induction at
$N=5$, since for $N=4$ all polynomials that one can write down are
zero (and hence cyclically reducible) because the $\eps$-factor
vanishes by symmetry and momentum conservation relations.

\subsubsection*{Variables, monomials, polynomials}

First, we define the following variables: \be
 s_{ij} = g_{\mu \nu} p_i^\mu p_j^\nu, \qquad 1 \leq i < j \leq N \ ,
\ee and \be
 \label{eq:epsdef}
 \eps_{ijkl} = \eps_{\mu \nu \rho \gs} p_i^\mu p_j^\nu p_k^\rho p_l^\gs, \qquad
 1 \leq i < j < k < l \leq N \ .
\ee Below, we will often encounter $s_{ij}$ and $\eps_{ijkl}$ where
the indices are not ordered in the above way; we take these
expressions to be defined in the obvious way: $s_{ji} = s_{ij},
\eps_{jikl} = - \eps_{ijkl}$, etc. 
Next, we define a set of monomials \be
 m_i, \qquad 1 \leq i \leq M_N \ ,
\ee consisting of products of one $\eps$- and one $s$-variable. Here
\be
 M_N = \binom{N}{2} \binom{N}{4} \ ,
\ee is the product of the number of $\eps$- and $s$-variables.
Finally, we define $\cP_N$ to be the $M_N$-dimensional vector space
generated by the $m_i$. We will thus represent polynomials $q=q^i
m_i$ as vectors $q^i$ consisting of $M_N$ numbers.

\subsubsection*{Relations}

Define the subspace $\cR_N$ to be the space generated by all
relations between the $m_i$. These relations arise as follows:
\begin{enumerate}
 \item
  Relations from momentum conservation on $s$-variables. These relations take the form
  \be
   \eps_{ijkl} \Bigg( \sum_{x \neq m} s_{m x} \Bigg) = 0 \ ,
  \ee
  for all values of $i,j,k,l,m$ with $i < j < k < l$.
 \item
  Relations from momentum conservation on $\eps$-variables. These relations take the form
  \be
   \Bigg( \sum_{x \not \in \{ i,j,k \} }\eps_{xijk} \Bigg) s_{lm} = 0 \ ,
  \ee
  for all values of $i,j,k,l,m$ with $i < j < k$ and $l < m$.
 \item
  Schouten identities: an antisymmetrized five-index tensor in four dimensions must vanish identically, meaning that
  \be
   \eps_{ijkl} s_{mn} + \eps_{jklm} s_{in} + \eps_{klmi} s_{jn} + \eps_{lmij} s_{kn} + \eps_{mijk} s_{ln} = 0 \ ,
  \ee
  for all values of $i,j,k,l,m,n$ with $i<j<k<l<m$.
\end{enumerate}
In this way, we find $R_N$ relations of the form ${r_\ga}^j m_j =
0$, where $\ga$ labels the different relations. We will think of $r$
as an $R_N \times M_N$ matrix in what follows. The relations will
not all be independent (that is, the rank of $r$ will be less than
$R_N$), but all that matters to us is that we have found a set that
spans $\cR_N$.

In what follows, it turns out to be useful to equip $\cP_N$ with a
metric, which we take to be the identity matrix in the basis
provided by the $m_i$. When we have a matrix such as $r$ whose rows
span a certain space, we will denote a matrix whose row vectors span
the perpendicular space by $(r^\perp)$: \be
 \forall \ga, \gb: \qquad (r^\perp)_{\ga l} {r_\gb}^l = 0 \ .
\ee This construction is implemented in Mathematica by the command
\verb@NullSpace@.

\subsubsection*{Operators}

Next, we construct two operators:
\begin{itemize}
 \item
  An $M_N \times M_N$ matrix ${c^i}_j$ that represents the cyclic permutations of the external particles on $\cP_N$. For example, $\eps_{1234} s_{12} \to \eps_{2345} s_{23}$ under a cyclic permutation, which can be written as $m_a \to m_b$ for some $a$ and $b$. Thus, ${c^j}_a = 1$ for $j=b$ and ${c^j}_a =0$ for $j\neq b$.
 \item
  An $M_{N-1} \times M_N$ matrix ${\pi^I}_j$ that projects onto the polynomials that survive when we set the momentum $p_N = 0$. In our implementation, we order the variables in such a way that the first $M_{N-1}$ monomials $m_i$ are the ones that do not involve $p_N$, so the matrix $\pi$ consists of an $M_{N-1} \times M_{N-1}$ identity matrix adjoined by $M_N - M_{N-1}$ columns of zeroes.
\end{itemize}

\subsubsection*{Cyclically reducible polynomials}

We will denote the space of all cyclically reducible polynomials by
$\cZ_N$. That is, $\cZ_N$ consists of all $q^j$ such that \be
 \label{eq:crpdef}
 {(1 - c)^i}_j \, q^j \in \cR_N, \qquad {\pi^I}_j q^j \in \cZ_{N-1} \ .
\ee The first equation imposes cyclic symmetry (up to relations);
the second one the reduction to a cyclically reducible polynomial of
one particle less. We rewrite the first equation as \be
 (r^\perp)_{\ga i} \; {(1 - c)^i}_j \; q^j = 0 \ .
\ee Thus, we are looking for all vectors perpendicular to the rows
of the matrix $r^\perp \cdot (1-c)$. 
Similarly, we can write the second equation in (\ref{eq:crpdef}) as
the search for all $q$ that satisfy \be
 z_{N-1}^\perp \cdot \pi \cdot q = 0 \ ,
\ee where $z_{N-1}$ is a matrix whose rows span $\cZ_{N-1}$. 
It follows that in order to construct $z_N^\perp$, all we need to do is combine the
rows of $r^\perp \cdot (1-c)$ and $z_{N-1}^\perp \cdot \pi$ into a
single matrix. We could of course construct $z_N$ from this by
looking for a basis of perpendicular vectors, but as we saw it is
$z_N^\perp$ which we will need at the next step of the induction, so
we do not need to do so.

Finally, if we want to know the number of cyclically reducible
polynomials up to relations, it is $M_N -
\rank\left(z_N^\perp\right)$. Therefore, the `true' number of
cyclically reducible polynomials (the ones that are not zero by
relations) is \be
 Z_N^{(\eps s)} = M_N - \rank\left(z_N^\perp\right) - R_N \ ,
\ee where we included a superscript $(\eps s)$ to remind the reader
that, in the example we treated in this section, we would still only
have calculated the number of basis polynomials containing terms of
the form $\eps s$. To complete the calculation, we then have to run
the analogous code for polynomials of the form $sss$, to find the
cardinality of the total basis: \be
 Z_N = Z_N^{(\eps s)} + Z_N^{(sss)} \ .
\ee For different degrees, we of course have similar expressions.

As a first check on our algorithms, we have checked that for scalar
variables, up to degree $8$, the analytic results of section
\ref{sec:scalarvariables} are reproduced.

\subsection{Results}
In this section, we collect the results of our computer
calculations, and comment on how much further this approach can be
taken.

\subsubsection*{The $(\ga')^2$ term}

For degree $d=4$, corresponding to the $(\ga')^2$ term in the string
expansion, the exact answer for the amplitude was found by
Stieberger and Taylor in \cite{Stieberger:2006te}. Therefore, this
is a good testing ground for our algorithms. We find that in this
case the number $Z_N$ of cyclically reducible polynomials for $N$
particles is \be
 Z_1=Z_2=Z_3=0, \qquad Z_4=2, \qquad Z_{N \geq 5} = 3 \ .
\ee To check that the number of cyclically reducible basis
polynomials indeed becomes constant, we have checked that $Z_N=3$
for $5 \leq N \leq 11$. It is worth noting that, contrary to the
scalar case, constancy does not set in at $N=d$ but at $N=d+1$. The
reason for this is that at $N=4$, the $\eps$-variables still vanish
due to momentum conservation.

An explicit $N$-independent realization of the three basis
polynomials is
\be P_1  =  [s_{12} s_{34}] \quad , \quad P_2  =  [s_{13} s_{24}] \quad , \quad
P_3  =  \sum_{i<j<k<l} \eps_{ijkl} \ .
\ee
In the first two polynomials, the square brackets mean that we sum
over all cyclically equivalent sets of indices, just like we did in
the scalar variable case. As we mentioned before, for the
$\eps_{ijkl}$ terms this cannot be done, which is why we use a
different summation.

We have checked that up to $11$ particles, the Stieberger-Taylor
answer can be expressed in terms of these polynomials as \be
 Q_2^{(N)} = -\frac{\pi^2}{12} \left( P_1 - P_2 + 4i P_3 \right)
 \ .
\ee This is a significant simplification compared to the general
formula given in equation (74) of \cite{Stieberger:2006te}. In fact,
the result can be written even more compactly as \be
 \label{eq:degfouranswerspin}
 Q_2^{(N)} = -\frac{\pi^2}{6} \sum_{i<j<k<l} \langle ij \rangle [jk] \langle kl \rangle [li] \ ,
\ee where we have used standard spinor notation.

\subsubsection*{The $(\ga')^3$ term}

After checking our algorithms with known results, we are now ready
to produce some new ones. The general answer for the $(\ga')^3$ term
in the MHV amplitude is not known in the literature. However, the
full amplitudes for $4$, $5$, $6$ \cite{Stieberger:2006te} and $7$
\cite{Stieberger:2007jv} external gluons is known from the work
of Stieberger and Taylor, so we can extract the $(\ga')^3$
coefficient from their expressions\footnote{In fact, since their
$7$-particle amplitude is written in terms of complicated Euler
integrals, it is quite difficult to extract an explicit $\al$
expansion from it, and we have not succeeded in doing so.
Fortunately, as we will see, for our purposes it is sufficient to
know the six-particle amplitude.}.

Using our computer program, we have found that the number $Z_N$ of
cyclically reducible basis polynomials up to relations for $d=6$ and
$N$ particles is \be
 Z_1 = Z_2 = Z_3 = 0, \qquad Z_4 = 2, \qquad Z_5 = 6, \qquad Z_{N \geq 6} = 13 \ .
\ee Note that here, contrary to the $(\ga')^2$ case, constancy of
$Z_N$ sets in at the `naive' value of $N=6$. We have checked the
constancy of $Z_N$ for up to $8$ particles and for the terms
involving $\eps$-tensors up to $9$ particles.

Since we have the full set of basis polynomials at $N=6$, we can
expand the answer of Stieberger and Taylor in this basis. We
conjecture that this in fact gives the general $(\ga')^3$ answer for
any $N$, as was the case for $(\ga')^2$. Because of the large number
of relations, there are many equivalent ways to write the answer;
one such way is \bea
 \label{eq:degsixanswer}
\!\!\! Q_3^{(N)} & = & \frac{\zeta(3)}{24} \Bigg( 42 [s_{12} s_{34} s_{56}] + 18 [s_{13} s_{24} s_{56}] - 9 [s_{13} s_{23} s_{56}] + 9 [s_{13} s_{25} s_{46}]
 -3 [s_{14} s_{25} s_{36}] \ret
 &&  + 36 [s_{12} s_{15} s_{34}] - [s_{12} s_{12} s_{12}] + 96i [\eps_{1234} s_{56}]
+24i [\eps_{1234} s_{45}] -24i [\eps_{1234} s_{35}] \Bigg) ,  \eea
where once again, the square bracket notation means that we should
sum over all index sets with the same cyclic ordering. Note that
this expression contains only ten terms; the coefficients of the
other three basis vectors are vanishing.

To check that this expression gives the answer for general $N$, we
have checked that also for 4 and 5 particles, it reproduces the
Stieberger-Taylor answer. It would be an even better check to see if
this answer also reproduced the seven-particle answer (where we have
13 coefficients to compare, instead of the 2 and 6 for $N=4,5$), but
due to the technicalities mentioned in the last footnote, we have
not been able to carry out this calculation. Instead, we will now
turn to a different and quite nontrivial check of
(\ref{eq:degsixanswer}): we will test if it behaves well under
collinear limits.

\subsubsection*{Collinear limits}

As we mentioned at the beginning of this section, the polynomial
$Q_3^{(N)}$ should satisfy \be
 Q_3^{(N)} \bigg|_{k^\mu_N = \ga p^\mu, p^\mu_{N-1} = (1-\ga) p^\mu} = Q_3^{(N-1)}
 \bigg|_{p^\mu_{N-1} = p^\mu} \ .
\ee Let us now check explicitly that this is the case. To this end,
we rewrite all terms in (\ref{eq:degsixanswer}) in the collinear
limit. For example, we have \bea
  \label{eq:collinearterm}
 {[}s_{12} s_{34} s_{56}] & \rightarrow & (s_{12} s_{34} s_{56}) \\
 && + \ga (s_{12} s_{34} s_{5x}) + \mbox{cyclic} \ret
 && + (1-\ga) (s_{12} s_{34} s_{5x}) + \mbox{cyclic} \ret
 && + (\ga - \ga^2) (s_{23} s_{4x} s_{x1}) + (\ga - \ga^2) (s_{4x} s_{x1} s_{23}) +
 (\ga - \ga^2) (s_{x1} s_{23} s_{4x}) \ . \nonumber
\eea Here, on the right hand side, all expressions are for $N-1$
particles. The index $x$ stands for the index with value $N-1$; the
round brackets mean that we sum over all index sets that have the
same cyclic structure as the term within brackets, where the $x$
always remains $N-1$. The first line on the right hand side comes
from the terms on the left hand side where no index equals $N$ or
$N-1$. The second line comes from the terms where one index equals
$N$ and no index equals $N-1$; the third line comes from the
converse. The fourth line comes from the terms where both an index
$N$ and an index $N-1$ are present. When adding the terms, we see
that the $\ga$-dependent terms in the first three lines cancel, and
the constant terms add up to the required term of one particle less.
The nontrivial part of the answer comes from the fourth line, and we
find \be
 {[}s_{12} s_{34} s_{56}]_{N} \rightarrow [s_{12} s_{34} s_{56}]_{N-1} + 3 (\ga - \ga^2) (s_{23} s_{1x} s_{4x})
 \ .
\ee Carrying out this procedure for every term in
(\ref{eq:degsixanswer}), we find \bea
 \label{eq:collinearlimit}
 Q_3^{(N)}&  \rightarrow & Q_3^{(N-1)} + \frac{\zeta(3)}{12} (\ga - \ga^2) \times \\
 && \Bigg\{ 18(s_{12}s_{1x}s_{3x}) + 18(s_{13}s_{2x}s_{4x}) + 6(s_{1x}s_{1x}s_{1x}) \ret
 && + 18(s_{23}s_{1x}s_{3x}) + 36(s_{23}s_{1x}s_{4x}) + 18(s_{24}s_{1x}s_{3x}) \ret
 && -12i(\eps_{x123}s_{2x}) +24i(\eps_{234x}s_{1x}) -24i(\eps_{123x}s_{3x}) -24i(\eps_{134x}s_{2x}) -48i(\eps_{123x}s_{4x}) \Bigg\} \ . \nonumber
\eea In principle, there could have been terms cubic in $\ga$ in
this expression, coming from terms where three $s$-indices become
$N$ or $N-1$, but it turns out that these terms cancel. Moreover,
the terms linear and quadratic in $\ga$ turn out to multiply the
same polynomial up to a sign. At first sight, however, this
polynomial does not seem to vanish. Fortunately, one can show using
momentum conservation and the Schouten identities, that it actually
{\em does} vanish. We give the details of this calculation in
appendix \ref{app:collinearity}. Inserting this result, we find that
\be
 Q_3^{(N)} \rightarrow Q_3^{(N-1)} \ ,
\ee so the answer (\ref{eq:degsixanswer}) indeed has the correct
behavior under collinear limits, as we wanted to show.

\subsubsection*{Order $(\ga')^4$ and beyond?}

Using our current Mathematica implementation, it seems difficult to
continue our computer calculations beyond $d=6$. For example, the
calculation for $d=6$ and $N=8$ takes roughly an hour on an average
desktop. On the other hand, using faster computers and a more
efficient programming language, it might be possible to carry out
the $(\ga')^4$ calculation up to 8 particles, which is the naive
number of particles for which we should find the full basis of
cyclically reducible polynomials.

However, this calculation will only give us the cardinality of the
basis for the space of cyclically reducible polynomials, which, if
the $2^d/d$ behavior we found in the scalar case is a good measure,
we expect to be of order 40 or so. This number is not too large, and
the construction of an actual basis should not be too difficult
either, but the problem is that after that, we do not have anything
to compare to: the eight-particle MHV amplitude has only been
calculated up to order $(\ga')^2$. Thus, it would not be possible to
express the known answer in our basis and find a conjecture for the
full $N$-independent answer.\footnote{Possibly, the techniques that
were developed quite recently in \cite{Berkovits:2008ic} could be
helpful in finding explicit expressions for higher-order corrections
to higher point amplitudes.} The best one could do at the moment
would be to use the seven-particle answer at $(\ga')^4$, which would
fix only part of the coefficients of the full answer.

Needless to say, these problems only get more severe at even higher
orders. Even if one could find smarter algorithms to construct a
basis for cyclically reducible polynomials of degree 10, we would
need at least the 10-particle amplitude up to order $(\ga')^5$ to
express the full answer in terms of this basis.

One solution to this problem might be to study polynomials that have
the correct reduction properties under collinear limits as well as
soft limits. In our $(\ga')^3$ calculation, we have seen that
requiring good collinear limits relates a polynomial of the form
$[s_{12} s_{34} s_{56}]$ to polynomials which have several indices
that are equal. The latter polynomials can be defined for a smaller
number of particles. Thus, using a basis of `collinear cyclically
reducible polynomials' of degree $d=8$, we might find that this
basis already reaches its maximum size for some number $N<8$ of
particles. If this is the case, the general $(\ga')^4$ answer can be
constructed using our methods above and the results that are known
in the literature.

However, also this approach will have a limited applicability -- it
appears unlikely that it will work for degrees above 4. Therefore,
it seems that ultimately, one should go back to the analytic
approach. In this respect, the answer (\ref{eq:degfouranswerspin})
gives hope due to its simplicity. It would be nice to express
(\ref{eq:degsixanswer}) in an equally simple way in spinor notation,
so one can find a pattern that might extend to higher orders.
Unfortunately, the most naive generalization, \be
 \sum_{i<j<k<l<m<n} \langle ij \rangle [jk] \langle kl \rangle [lm] \langle mn \rangle [ni] \ ,
\ee though it turns out to be cyclically reducible, cannot be the
answer we are looking for: it vanishes after a single soft limit. We
have found some expressions for $Q_3^{(N)}$ in terms of spinor
notations that do work, but they are not any nicer than
(\ref{eq:degsixanswer}), and do not provide any intuition on how to
proceed to higher degree.

\section{CSW rules for the Abelian DBI action}
\label{sec:action} The previous section focused on solving the
constraints on the gluon scattering amplitude of the superstring
which follow from supersymmetry and kinematics. As emphasized
before, one of the reasons one is interested in these amplitudes is
to derive an expression for the string effective action from them.
This effective action can then be taken off-shell to further study
the behavior of the string theory. However, the effective action may
be obtained by other methods, and given an expression for the
effective action, one may conversely ask what amplitudes this action
generates. Especially in the case of amplitudes of strings ending on
a single D-brane quite a lot is known about the effective action, in
particular because in this case one has a reliable derivative
expansion. This leads for instance to the well-known
Dirac-Born-Infeld action, see e.g.\ \cite{Tseytlin:1999dj} to which
the reader is also referred for general background information.
Motivated by a desire to obtain CSW-style perturbation theory, we
study in this section the Abelian Dirac-Born-Infeld effective action
reduced to four dimensions and its tree level scattering amplitudes.

\subsection{Dirac-Born-Infeld in four dimensions}
In general, the string effective action for Abelian fields can have
terms of the form
\begin{equation}\label{eq:dimanaDBI}
\mathcal{L}_{\textrm{deriv}} \sim \al^m \partial^{n} F^p \ .
\end{equation}
Simple dimensional analysis restricts $p = (m +2 -n/2)$, and it is
also known that $p$ cannot be odd because of worldsheet parity (we
will shortly encounter a second reason why this is true in four
dimensions). The corrections to ordinary electrodynamics are known
at the level of the action for
\begin{enumerate}
\item terms without derivatives ($n=0$)
\item terms with four fields \cite{deRoo:2003xv} ($p=4$)
\item (up to) four derivative terms \cite{Wyllard:2000qe} ($n=4$)
\end{enumerate}
The leading derivative corrections are given by the DBI action,
while the second simply follow from the known Veneziano amplitude
\eqref{eq:fourpoints} for four photons. Although interesting, we
will not consider the third series of terms in this paper. The DBI
action reads
\begin{equation}\label{eq:DBIaction}
S_{\textrm{DBI}} = -1 + \frac{1}{\pi^2 g_s \al^5}\int d^{10}x
\sqrt{-\det\left(\eta_{\mu \nu} + \pi \al F_{\mu \nu} \right)} \ ,
\end{equation}
which can be derived in several ways \cite{Tseytlin:1999dj}. In this
section, we will put $g_s=1$ and drop the irrelevant `$-1$'
term\footnote{The explicit dependence on $\pi$ will be useful to
make some comments about the `transcendentality' of the scattering
amplitudes.}. We will
adhere to the convention that
\begin{equation}
Z[J] = \int e^{i S } \ ,
\end{equation}
but note that in the $n$-point amplitudes given below we omit an overall normalization factor of $i$.

Although the above (part of the) effective action can be studied in
any number of dimensions in principle, here only the dimensional
reduction to four dimensions will be considered. In other words, in
terms of the scattering amplitudes $4$ directions will be selected
and all momenta and polarization vectors of the external fields will
be embedded in these dimensions. With this choice it is obvious that
in practice one can simply ignore all off-dimensional parts of the
action and replace $10 \rightarrow 4$ by ordinary dimensional
reduction. Since in four dimensions it is known that the determinant
in (\ref{eq:DBIaction}) is a Lorentz invariant of maximal order $4$
in the field strength, one can write down a generic ansatz for the
determinant in terms of these. Both the ansatz as well as the
determinant can be evaluated explicitly by simply writing
$\eta_{\mu\nu}$ and $F_{\mu\nu}$ as matrices\footnote{We define
$\tilde{F}_{\mu\nu} \equiv \frac{\ii}{2}
\epsilon_{\mu\nu}^{\rho\sigma} F_{\rho \sigma}$}. Comparing the
expressions gives after some experimentation
\begin{equation}
-\det\left(\eta_{\mu \nu} + \pi \al F_{\mu \nu} \right) = 1 +
\frac{\pi^2 \al^2}{2} F_{\mu \nu} F^{\mu\nu} + \frac{\pi^4
\al^4}{16} \left(F_{\mu \nu} \tilde{F}^{\mu\nu} \right)^2 \ .
\end{equation}

In the following we will find it useful to write the DBI action in
terms of selfdual and anti-selfdual field strengths. In four
dimensions the field strength tensor can be decomposed uniquely in
terms of a selfdual and an anti-selfdual field strength,
\begin{equation}
F_{\mu \nu} = F^{+}_{\mu \nu} + F^{-}_{\mu \nu} \ ,
\end{equation}
which in turn have a natural expression in terms of spinor
variables,
\begin{equation}
F_{\mu \nu} \leftrightarrow F_{\alpha \beta \dalpha \dbeta} =
\epsilon_{\alpha \beta} F^+_{\dalpha \dbeta} + \epsilon_{\dalpha
\dbeta} F^-_{\alpha \beta} \ .
\end{equation}
Note that in our conventions,
\begin{equation}
\tilde{F}_{\alpha \beta \dalpha \dbeta} = \epsilon_{\alpha \beta}
F^+_{\dalpha \dbeta} - \epsilon_{\dalpha \dbeta} F^-_{\alpha \beta}
\ .
\end{equation}
This is of course nothing but the observation that the tensor
$F_{\alpha \beta \dalpha \dbeta}$ must be antisymmetric under
exchange of $(\alpha \dalpha) \leftrightarrow (\beta \dbeta)$, so it
can be decomposed in terms that are anti-symmetric in either the
dotted or undotted indices. As there is only one antisymmetric 2
$\times$ 2 matrix, these terms must be proportional to the 2
$\times$ 2 $\epsilon$ symbol, and the above conclusion follows for a
choice of normalization contained in the equation. In terms of the
spinor field strengths the action \eqref{eq:DBIaction} reads
\begin{equation}\label{eq:DBIselfdual}
S_{\textrm{DBI}} = \frac{1}{\pi^2 \al^2}\int d^{4}x
\sqrt{\left(1+\frac{\pi^2 \al^2}{8} (F_+^2 + F_-^2) \right)^2
-\frac{\pi^4 \al^4}{16} F_+^2 F_-^2 } \ .
\end{equation}
This can now be expanded in powers of $\al$, either symmetrically in
the field strengths
\begin{align}\label{eq:biexpsd}
\mathcal{L}_{\textrm{DBI}} = \frac{1}{4} F_+^2 - \frac{\pi^2 \al^2}{32} \left(F_-^{2} F_+^{2}\right) + & \frac{\pi^4 \al^4}{256}  \left(F_-^{4} F_+^{2} + F_+^{4} F_-^{2}\right)  \nonumber \\
 & - \frac{\pi^6 \al^6}{2048} \left(F_-^{6} F_+^{2} + 3 F_-^{4} F_+^{4} + F_+^{6} F_-^{2}\right) + \mathcal{O}(\al^8)
 \ ,
\end{align}
or asymmetrically
\begin{equation}\label{eq:MHVgenact}
\mathcal{L}_{\textrm{DBI}} = \frac{1}{4}F_+^2 - \frac{ \pi^2
\al^2}{32} \left(\frac{F_-^{2} F_+^{2}}{1+ \frac{1}{8} \pi^2 \al^2
F_-^{2} }\right) + \frac{ \pi^4 \al^4}{256} \left(\frac{F_-^{2}
(F_+^{2})^2}{(1+ \frac{1}{8} \pi^2 \al^2 F_-^{2})^3 }\right) +
\mathcal{O}\left(F_+^{6}\right) \ ,
\end{equation}
where we have furthermore added the topological density ($
\frac{1}{4} F_{\mu \nu} \tilde F^{\mu \nu} = \frac{1}{8} (F_+^2 - F_-^2)$). As we will see below, the last form
(\ref{eq:MHVgenact}) is more geared towards deriving MHV amplitudes.

\subsection{Deriving scattering amplitudes from the DBI action \label{sec:derDBI} }
Writing the DBI action in terms of selfdual and anti-selfdual fields
is advantageous because for \emph{on-shell} photon fields of
definite helicity, $A^{+}$ and $A^-$, these field strengths
simplify. The polarization vectors of these on-shell fields are
given in our conventions\footnote{Our spinor conventions are such
that $p_{\alpha \dalpha} = \sigma^{\mu}_{\alpha \dalpha} p_{\mu}$,
from which $g_{\mu \nu} \leftrightarrow 2 \epsilon_{\alpha \beta}
\epsilon_{\dalpha \dbeta}$. Dotted spinors are called holomorphic.
Furthermore $\epsilon_{\alpha \beta} \epsilon^{\beta \gamma} =
\delta_{\alpha}^{\gamma}$, and we have $\braket{ab} = a^{\dalpha}
b_{\dalpha} = a^{\dalpha} b^{\dbeta} \epsilon_{\dbeta \dalpha}$,
$\sbraket{ab} = a_{\alpha} b^{\alpha} = a^{\alpha} b^{\beta}
\epsilon_{\alpha \beta}$. These conventions largely follow
\cite{Boels:2008ef}.} by
\begin{equation}
\epsilon^+_{\alpha \dalpha} = \sqrt{2} \frac{\eta_{\alpha}
p_{\dalpha}}{\sbraket{\eta p}} \: , \quad \epsilon^-_{\alpha
\dalpha} = \sqrt{2} \frac{\eta_{\dalpha} p_{\alpha}}{\braket{\eta
p}} \ .
\end{equation}
Therefore we have in spinor notation
\begin{eqnarray}\label{eq:extlinefact}
F_{\dalpha \dbeta}^{+}[A^+] =  i \sqrt{2} p_{\dalpha} p_{\dbeta} \:
,
& & F_{\dalpha \dbeta}^{+}[A^-] = 0 \nonumber \\
F_{\alpha \beta}^{-}[A^-] = i \sqrt{2} p_{\alpha} p_{\beta} \: , & &
F_{\alpha \beta}^{-}[A^+] = 0 \ .
\end{eqnarray}
Note that the $(+,-)$ superscripts on the fields strengths only
correspond to the helicity quantum number \emph{on-shell}. This will
be very important when considering larger Feynman diagrams. The
above equations imply that in a given Feynman diagram derived from
the four-dimensional DBI action the contractions between vertices
and external states are in fact very simple. For these `external
line' factors field strength superscripts immediately translate into
helicity superscripts. For the four-point scattering amplitude for
instance, which is controlled by just one vertex from
\eqref{eq:biexpsd}, it follows therefore immediately that the
amplitude derived from the DBI action reads
\begin{equation}
A^\textrm{DBI}_4(1^+ 2^+ 3^- 4^-) = - \frac{\pi^2
\al^2}{2}\braket{1 2}^2 \sbraket{3 4}^2 \ ,  \label{eq:MHV-4-point}
\end{equation}
and that the following amplitudes vanish
\begin{equation}
A^\textrm{DBI}_4(1^+ 2^+ 3^+ 4^+) = A^\textrm{DBI}_4(1^- 2^- 3^- 4^-) = A^\textrm{DBI}_4(1^- 2^+ 3^+ 4^+) = A^\textrm{DBI}_4(1^+ 2^- 3^- 4^-) = 0 \ .
\end{equation}
The `DBI' superscript is written here and elsewhere in this section to remind the reader that
since these amplitudes were derived from the DBI action, they are
only the lowest order correction in $\al$ to the full superstring
amplitudes. Note that the MHV amplitude in (\ref{eq:MHV-4-point}) is
the full amplitude, so it does not include a suppressed sum over
permutations of external particles.

For more particles we need to take into account the `interior' of
the Feynman graphs: contractions between the photon fields within
the field strength tensors in the different vertices. Note that for
this the $(+,-)$ superscripts do not correspond to the helicity
quantum number. As can be checked explicitly in Feynman-'t Hooft
gauge for instance, these contractions are \bea
\ldots F_{\dalpha \dbeta}^+: :F_{\alpha \beta}^-\ldots & = & -i \left(\frac{p_{\alpha \dalpha} p_{\beta \dbeta} + p_{\beta \dalpha} p_{\alpha \dbeta}}{p^2} \right) \nonumber \\
\ldots F_{\dalpha \dbeta}^+: :F_{\ddelta \dgamma}^+\ldots & = &  i \left(\epsilon_{\dalpha \ddelta} \epsilon_{\dbeta \dgamma} + \epsilon_{\dbeta \ddelta} \epsilon_{\dalpha \dgamma} \right) \nonumber \\
\ldots F_{\alpha \beta}^-: :F_{\delta \gamma}^-\ldots & = &  i
\left(\epsilon_{\alpha \delta} \epsilon_{\beta \gamma} +
\epsilon_{\alpha \gamma} \epsilon_{\beta \delta} \right) \ .
\label{eq:ymcontract} \eea Hence, the contraction between photon
field strengths of the same type is nonzero and \emph{local} in
space-time as it simply does not depend on the momentum. This is the
result of a cancellation of the propagator pole through a numerator
factor. It is only non-local between field strength tensors of
opposite type. Note that this `local' property of the contraction
does not depend on any on-shell conditions: the cancellation occurs
off-shell. A similar phenomenon was discussed in the context of
effective Higgs-gluon couplings through a top quark loop in
\cite{Boels:2008ef}; we will come back to this below.

Using the contractions in eqs. \eqref{eq:ymcontract}, we can now
calculate 6-gluon amplitudes up to $\mathcal{O}(\alpha'^5)$ from
\eqref{eq:biexpsd}. First of all, it is easy to see from that
equation that several amplitudes vanish,
\begin{equation}\label{eq:sixpointvanishing}
A^\textrm{DBI}_6(++++++) = A^\textrm{DBI}_6(+++++-) =
A^\textrm{DBI}_6(-----+) = A^\textrm{DBI}_6(------)=0  \ .
\end{equation}
This follows because there are simply no vertices with six self-dual
field strengths or five of them and one anti-selfdual field
strength. Therefore, there are no diagrams at tree level. This
brings us to the MHV amplitude $A_6(++----)$ and its conjugate.
There are only two contributions: a direct one $ i (\pi \al)^4/2^8$
from the vertex in
the DBI action (\ref{eq:biexpsd}) with six fields, and a
contribution which consists of two four-point vertices. In the
latter case, two selfdual field strengths need to be contracted. It
is easy to see that this yields a contribution exactly proportional
to the six-point vertex in the action. In fact, in terms of Wick
contractions the relevant part of the calculation is the contraction
between the two vertices:
\begin{equation}
 \frac{(i)^2}{2} \frac{\pi^4 \al^4}{2^{10}}\langle :\textrm{ext. fields}:
:F_+^2 F_-^2: :F_+^2 F_-^2:\rangle _0  \ ,
\end{equation}
where the factor of a $\frac{1}{2}$ comes from the exponential of
the action. For a tree level contribution to the amplitude under
study, two $F_+$ have to be contracted between the two vertices.
This can be done in $4$ different ways. In addition, the
`propagator' from \eqref{eq:ymcontract} gives another factor of $2i$
after performing the symmetrization. Hence the total prefactor is
$-i(\pi \al)^4 /2^8$ which is the opposite of that of the direct vertex,
yielding  an exact cancellation off-shell.
Therefore, the MHV amplitude is given by
\begin{equation}
\label{eq:6mhv}
A^\textrm{DBI}_6(1^+ 2^+ 3^- 4^- 5^- 6^-) = 0 \ .
\end{equation}
The conjugate $\overline{\textrm{MHV}}$ amplitude follows by conjugation and
hence also vanishes. This is in line with \cite{Rosly:2002jt}, who
argue that only helicity conserving amplitudes are non-zero. Their
argument is based on the observation that since the classical
equations of motion have a $U(1)$ S-duality symmetry, this symmetry
should lead to selection rules. Since the selfdual and anti-selfdual
solution have charge $+1$ and $-1$ under this symmetry respectively,
and are in fact solutions to the full field equations\footnote{Note
that \eqref{eq:DBIselfdual} implies that a variation of the action
is proportional to $F^+ F^-$, up to the electromagnetic $F^2$
leading term. Hence any self-dual or anti-self-dual field is a
solution to the field equations. }, conservation of the full $U(1)$
implies that only helicity conserving amplitudes are non-zero. This
argument is somewhat unusual in that one uses nonperturbative
symmetries to prove perturbative results. On the other hand, there
are instances in string theory where duality symmetries like
S-duality fix parts of the effective action. This in turn fixes the
perturbatively derived tree level scattering amplitudes
\cite{Green:1997tv}. Regardless of this reasoning, below, we will verify
the result explicitly by a diagrammatic argument.

The remaining helicity conserving $6$-point NMHV amplitude,
$A^\textrm{DBI}_6(+++---)$, is slightly more complicated as it will
involve one propagator. The result is
\begin{equation}\label{eq:6pointDBIampl}
A^\textrm{DBI}_6(1^+ 2^+ 3^+ 4^- 5^- 6^-) = - \frac{\pi^4 \al^4}{2^7}
\left( \frac{\sbraket{1 2}^2 \braket{5 6}^2 ([4|(1+2)|3\rangle)^2
}{(p_1 + p_2 + p_4)^2} \: + \: {\textrm{permutations}} \right) \ .
\end{equation}
Here, the sum is over all permutations that separately permute the
positive-helicity indices $(123)$ and the negative helicity-ones
$(456)$, but do not mix them. We also use the standard notation \be
 [i|j|k \rangle = (p^i)^\alpha \, p^j_{\alpha \dalpha} \, (p^k)^{\dalpha}.
\ee Note that this form of the six-point amplitude can be
interpreted as just a sum over all the factorization channels: it is
easily seen that the residues are the usual four-point functions.

\subsection{Helicity conserving amplitudes from DBI}
After these examples we can proceed to more general cases, but
before doing this let us make some simple kinematic observations.
The vanishing amplitudes in \eqref{eq:sixpointvanishing} easily
generalize to all multiplicity amplitudes, since there are simply no
diagrams with the required field strengths on the external lines.
This is of course in line with the target space supersymmetric Ward
identities which follow from the underlying supersymmetric string
theory discussed previously. Furthermore, note that the MHV
amplitude is at the very least required to be a `local' quantity:
this quantity cannot have \emph{any} kinematic poles. Therefore, it
must be proportional to the amplitude derived from the vertex with
the same number of field strengths as the number of particles in the
amplitude, just as was shown above in the case of the six-point
amplitude. Actually, the full diagrammatic calculation of an $n$-point MHV amplitude from the DBI action is structurally the same as
that calculation. Hence the MHV amplitude is expected to be
\begin{equation}
A^\textrm{DBI}_{n = {\rm even}} (1^+ 2^+ 3^- \ldots n^-) \stackrel{{\bf \large ?}}
     {\sim} (\al)^{n-2} \! \! \! \sum_{\textrm{perm}(3,\ldots,n) \notin
\textrm{cyclic}} \sbraket{12}^2 \braket{34}^2 \braket{56}^2 \ldots
\braket{n-1,n}^2 \, ,
\end{equation}
and only the proportionality constant needs to be calculated. Based
on the six-particle result \eqref{eq:6mhv}, one should be starting to suspect that
this is actually zero to leading order in $\al$. For sub-leading
orders generated by `beyond DBI' (higher derivative) contributions
to the effective action, the above form may still be relevant.

\subsubsection*{Vanishing of MHV amplitudes beyond the four-point function (and NMHV beyond six)}
One can show that some of the effects of (\ref{eq:ymcontract}) are
resummed when one employs a first order formalism. To derive this
one first introduces a first order action for the free
electromagnetic field pioneered for different purposes by Chalmers
and Siegel~\cite{Chalmers:1996rq},
\begin{equation}\label{eq:chalsiegaction}
S_{\textrm{CS}} =  \int d^4 x  \left[ \frac{1}{2}  C_{\dalpha \dbeta}
F^{\dalpha \dbeta}_+ - \frac{1}{4} C_{\dalpha \dbeta} C^{\dalpha
\dbeta} \right]  \ .
\end{equation}
Here $C$ is an auxiliary self-dual two-form. Integrating this out
from the above action gives back standard electrodynamics (and a
topological term). One can derive the propagators for the above
action. In a form similar to \eqref{eq:ymcontract} these read, \bea
\ldots C_{\dalpha \dbeta}: :F_{\alpha \beta}^-\ldots & = & i \left(\frac{p_{\alpha \dalpha} p_{\beta \dbeta} + p_{\beta \dalpha} p_{\alpha \dbeta}}{p^2} \right) \nonumber \\
\ldots C_{\dalpha \dbeta}: :C_{\ddelta \dgamma}\ldots & = & 0 \nonumber \\
\ldots F_{\alpha \beta}^-: :F_{\delta \gamma}^-\ldots & = & -  i
\left(\epsilon_{\alpha \delta} \epsilon_{\beta \gamma} +
\epsilon_{\alpha \gamma} \epsilon_{\beta \delta} \right)
\label{eq:ymcontractII} \eea The second line above
is the most important one: in this first order form contractions
between $C$ fields (which are $F_+$ on-shell) are trivial. Inspired by
\eqref{eq:chalsiegaction}, it is easy to see that in order
to obtain (\ref{eq:MHVgenact}) one needs to integrate out $C$ from
\begin{equation}\label{eq:abelacttolift}
S =   \int d^4 x \left[ \frac{1}{2}  C_{\dalpha \dbeta} F^{\dalpha
\dbeta}_+ - \frac{1}{4} (1 + \al^2 \pi^2 \frac{1}{8} F_-^2) C_{\dalpha \dbeta}
C^{\dalpha \dbeta} + \mathcal{O}(F_+^4) \right] \ .
\end{equation}
This form makes it obvious that the $4$-point MHV amplitude is the
only MHV amplitude: the perturbation series works analogous to the
case considered above, except for the second line in
\eqref{eq:ymcontractII} and the absence of vertices with one $F_+^2$
and multiple $F_-^2$ in the action. It is easy to see that this
action only leads to $4$-point MHV and $6$-point NMHV amplitudes.
More precisely,
\begin{equation}
A^{\textrm{DBI}}(++-^i) = A^{\textrm{DBI}}(+++ -^{i+1}) =0 \quad
\quad \forall \, i>2 \ .
\end{equation}
In principle this argument can be extended to NNMHV and beyond, but below we
follow a more elegant and parity-symmetric route.

This form of the action is notable for being easily lifted to
twistor space. Just as the Poincar\'e group acts linearly on
$\mathbb{R}^4$, the complexification of the conformal group of this
four-dimensional space acts linearly on $\CP^3$, which is known as
its twistor space. Given an action on space-time with spin $0,
\frac{1}{2},1$ fields, it is possible to construct an action on
twistor space which reproduces the perturbation series of the
original action, see for instance the review in \cite{Boels:2008ef}.
The resulting twistor action generates exactly one MHV amplitude:
the four-point one, while the rest vanishes. Unfortunately, the
higher-order terms seem to defy a neat formulation on the twistor
space. Also, the parity-symmetric structure of the amplitudes
uncovered below suggests that one should be looking for ambi-twistor
space action constructions \cite{Mason:2005kn} which are not nearly
as well understood.

\subsubsection*{Helicity conservation}
In the following we will find it useful to consider not the DBI
action in the form \eqref{eq:DBIaction}, but written using two
auxiliary complex scalar fields \cite{Rocek:1997hi}:
\begin{equation}
 \mathcal{L}_{\textrm{DBI}} = - \frac{1}{4} F_+^2 +
 \frac{i}{2 \pi \al} \left(-i a \bar{a} +   \lambda a -  \bar{\lambda} \bar{a} +
 \frac{\sqrt{\pi \al}}{2} a \bar{a} (\lambda -\bar{\lambda})  \right)
 - i \frac{\sqrt{\pi \al} \lambda }{8} F_+^2 +  i \frac{\sqrt{\pi \al} \bar{\lambda}}{8} F_-^2\ ,
 \label{eq:scalarDBI}
\end{equation}
Shifting the field $\lambda$ by $-i / \sqrt{\pi \al}$ and integrating
out the $\lambda$ and $a$ fields yields back \eqref{eq:DBIaction}.
The main reason this is an easier action to consider is that the
coupling of the field strengths is only through scalars. In fact,
this exact same coupling is considered in the literature for
non-Abelian fields for effective Higgs-gluon couplings through a top
quark loop \cite{Wilczek:1977zn}. It is not hard to see that since
the contractions between like-selfduality field strengths happen
off-shell, one can resum these effects into new effective vertices.
It is then important to calculate the correct constants multiplying
these effective vertices. These can be read of from
\cite{Boels:2008ef} by specializing to an Abelian gauge
group\footnote{Using the correct normalization of the color matrices
there.}. More specifically, in our conventions one can replace
\begin{equation}\label{eq:ymeffecrepl}
 -i \frac{\sqrt{\pi \al} \lambda}{8} F_+^2 \rightarrow -\frac{1}{4}
\frac{i \sqrt{\pi \al} \lambda}{2 + i \sqrt{\pi \al} \lambda} :F_+^2: \quad
i \frac{\sqrt{\pi \al} \bar{\lambda}}{8} F_-^2 \rightarrow \frac{1}{4}
\frac{i \sqrt{\pi \al} \bar{\lambda}}{2 - i \sqrt{\pi \al} \bar{\lambda}}
:F_-^2: \ ,
\end{equation}
as long as one keeps in mind that contractions between
like-selfduality field strengths are not allowed any more; only the
non-local contraction between $F_-$ and $F_+$ is still allowed.

Now we turn to the scalar part of \eqref{eq:scalarDBI}. The fields
$a$ and $\bar{a}$ can be integrated out exactly, which yields the
following polynomial scalar Lagrangian
\begin{equation}\label{eq:scalarpartofDBI}
\mathcal{L}_{\textrm{scalar}} = - \frac{1}{\pi \al} \left( \frac{\lambda
\bar{\lambda}}{2  + \sqrt{\pi \al} i (\lambda - \bar{\lambda})} \right) \ .
\end{equation}
Given the couplings in \eqref{eq:ymeffecrepl}, it is natural to
define new fields $k$ and $\bar{k}$ as
\begin{equation}\label{eq:defkfields}
k =   \frac{  \lambda}{2 + i \sqrt{\pi \al}\lambda} \quad \bar{k} =
\frac{ \bar{\lambda}}{2 - i
\sqrt{\pi \al} \bar{\lambda}}  \ .
\end{equation}
Plugging this field redefinition into the scalar part
\eqref{eq:scalarpartofDBI}, one obtains
\begin{equation}\label{eq:effectiveDBIaction4D}
\mathcal{L}_{\textrm{DBI}} = -\frac{1}{\pi \al} \left( \frac{2 k
\bar{k}}{1 - \pi \al k \bar{k}} \right) - \frac{1}{8} i \sqrt{\pi \al} k :F_+^2: +
\frac{1}{8} i \sqrt{\pi \al} \bar{k} :F_-^2: \ .
\end{equation}
This form of the DBI action is very convenient for deriving tree
level helicity amplitudes. Note that its
Feynman rules are such that only equal numbers of selfdual and
anti-selfdual field strengths appear on the external legs. This is a
direct diagrammatic verification of the claim (see also section \ref{sec:derDBI})
in \cite{Rosly:2002jt}: amplitudes derived from the DBI action conserve
helicity.

Loop level amplitudes are ill-defined for this action in general, not taking into account the
fact that in general loop calculations in an effective action only
estimate the effects of the UV theory. However, knowing that
amplitudes vanish at tree level implies that several loop level
amplitudes are rational functions of the external momenta as they
cannot have branch cuts. This seems to imply some of the annulus diagrams might also be simpler than expected. Note that most of the Jacobians encountered in the field transformations above will vanish in dimensional
regularization as they involve $\int d^dp \; 1 =0$.

\begin{figure}[t]
  \begin{center}
  \includegraphics[width=0.9\textwidth]{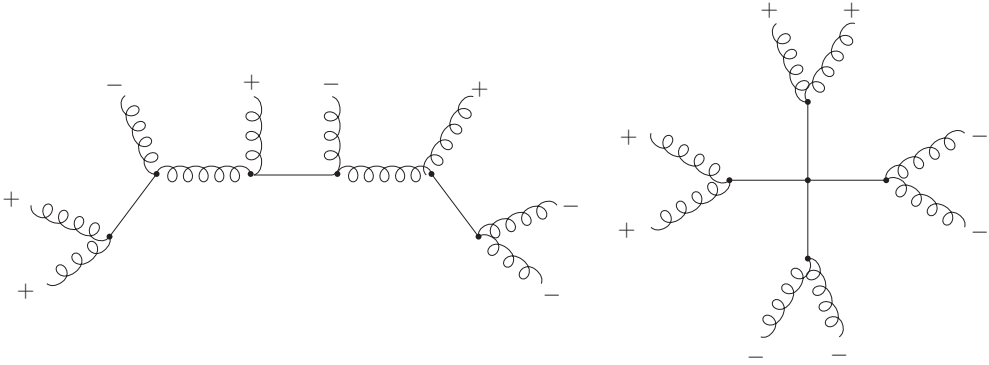}
  \caption{Illustration of the two types of contribution to the $8$-point amplitude:
  non-local (left) and local (right).}
  \label{fig:contribsto8point}
  \end{center}
\end{figure}

\subsection{Interpretation as (partial) CSW rules}
The non-zero amplitudes also follow from this action. As a
cross-check, the six-point amplitude \eqref{eq:6pointDBIampl} calculated above is seen to be
reproduced (the form of the diagrams is illustrated for different
purposes in figure \ref{fig:goodbadtermsrecursion}, see
appendix \ref{app:onshellrecDBI}). For the eight-point amplitude there
are only two types of contribution illustrated
in figure \ref{fig:contribsto8point}. The type displayed on the left
in that figure is the generalization of \eqref{eq:6pointDBIampl}, which can
again be interpreted as a sum over factorization channels. The other
is a direct contribution from the action generated by the helicity
conserving four-point scalar vertex in
\eqref{eq:effectiveDBIaction4D}
\begin{equation}\label{eq:localcontribto8point}
A^\textrm{DBI}_8(++++----)_{\textrm{local}} \sim  \pi^6 \al^6
\sum_{\textrm{perm}(1234)(5678)} \braket{1 2}^2 \braket{3 4}^2
\sbraket{5 6}^2 \sbraket{7 8}^2 \ .
\end{equation}
This also illustrates to what extent the DBI action generates
`CSW'-style perturbation theory rules. For our purposes, this can be
phrased as the question if given all $\textrm{N}^{i}$MHV amplitudes
(not vertices) one can calculate the $\textrm{N}^{i+1}$MHV
amplitudes. This question is however rather simple considering the
perturbation theory derived from \eqref{eq:effectiveDBIaction4D}
(which is already quite close) since there are not that many
non-zero contributions.

Given the MHV amplitude the NMHV amplitude can be calculated. For
this one uses the MHV amplitudes as vertices and recognizes that the
spinor products of the leg which needs to be taken off-shell exactly
correspond to the product of the Lorentz momentum flowing through
the leg. One replaces the Lorentz momentum by the off-shell Lorentz
momentum and arrives at the same expression as generated by the
action \eqref{eq:effectiveDBIaction4D}. Unfortunately this process
does not generate the complete NNMHV amplitude because of the
`local' contribution calculated in  \eqref{eq:localcontribto8point}.
The extra contributions are known though, since for
\begin{equation}
n= 4m \quad, \;  m > 1 \ ,
\end{equation}
fields all these can be calculated explicitly from combining four
scalar vertices leading to the natural generalization of
\eqref{eq:localcontribto8point},
\begin{equation}\label{eq:localcontribto8pointgen}
A^\textrm{DBI}_{2 i}\left((+)^i\, (-)^i\right)_{\textrm{local}} \sim
(\pi \al)^{2(i-1) } \!\! \!\! \!\!\sum_{\textrm{perm}(1\ldots i)(i+1
\ldots 2 i)} \! \braket{1 2}^2 \ldots \braket{i-1, i}^2
\sbraket{i+1, i+2}^2 \ldots \sbraket{2i-1,  2i}^2 \ ,
\end{equation}
where $i=2m$.
For example for twelve particles these come from diagrams with two
four-point vertices, etc. On the other hand for
\begin{equation}
n= 4m +2 \quad, \;  m \geq 1 \ ,
\end{equation}
the Dirac-Born-Infeld action does obey CSW-style Feynman rules
directly. The more important observation though is that since so
many amplitudes vanish, the perturbation theory is surprisingly
trivial.

\subsubsection*{More general amplitudes}
Applying the usual supersymmetric Ward identities to the derived
scattering amplitude formulae in this section will generate certain
amplitudes with specific combinations of quarks and scalars. These
amplitudes are generated from the supersymmetric completion of the
DBI action as written above. Note that since the supersymmetric
partners of the photon transform in the adjoint of the gauge group,
one obtains uncharged scalars and fermions and one therefore expects
no extra contributions at order $\al^0$ from these particular terms.
There are more terms however, which arise for scalars for instance
directly under the `square root' sign by dimensional reduction from
the ten-dimensional DBI action. It would be interesting to study
their coupling further.

\subsubsection*{Higher orders in $\al$}
As noted in the introduction to this section, several string
corrections to the DBI action are known. These are basically the
series of term related to the Veneziano amplitude which have four
fields and those terms in the effective action which have $4$
derivatives. Although both of these are interesting we will only
make some preliminary remarks about both of them.

\subsubsection*{Further remarks}
In principle, since the expression in \eqref{eq:scalarDBI} is highly
reminiscent of open string field theory one might suspect that one
can calculate also the above mentioned corrections with an action of
the same type. The starting point is writing a kinetic term for the
scalars, after which counter terms must be added to obtain the
correct correction to the Veneziano amplitude (which starts at order
$\al^4$). These terms may involve massive spin fields with spin to
mimic the states of the string. The problem is that although
constructing an extension of \eqref{eq:scalarDBI} which reproduces
the Veneziano amplitude along these lines is possible in principle,
in practice this procedure is fairly arbitrary. Note that related
observations play a major role in appendix
\ref{app:onshellrecDBI}, which discusses on-shell recursion for
the DBI action.

More promising, one can construct a similar action as in \eqref{eq:scalarDBI}
which reproduces the non-Abelian DBI action up to $\al^2$. In this respect it is
interesting to note that although there is no gauge invariant
derivative expansion, there is a gauge invariant expansion in terms
of angular momentum of the modes of the massive states.

\section{BCFW recursion relations in string theory}
\label{sec:recursion}
In this section we initiate the study of on-shell recursion relations of the BCFW type for string theory amplitudes. First we review the general derivation of on-shell recursion relations, followed by some general comments. After this we study some examples for which everything can be checked explicitly. Note that this section would do well in any seventies-string-theory-buzzword-bingo game as it involves concepts like duality, Regge behavior and resonance poles.

Before continuing let us note that at the very least the $\al^2$ correction obeys \emph{some} recursion relation: by it's close relation to the one loop all-plus gluon amplitude in pure Yang-Mills theory, it will obey the same recursion relations as studied in \cite{Bern:2005hs}. However, these were derived `experimentally' by studying the residue at infinity for a given BCFW-like shift of the known expression. It would be interesting to see if a similar procedure would work for our $\al^3$ result in section \ref{sec:alphaprime3}. For the rest of this section, we focus on BCFW's original idea.

\subsection{The main BCFW idea}\label{sec:genericdiscussion}
Following BCFW's idea \cite{Britto:2005fq}
quite closely, the momenta of $2$ external
particles which can be massive or massless in $D\geq 4$ dimensions
are shifted by a vector $n_{\mu}$
\begin{align}
p^{\mu}_i &\rightarrow \hat{p}^{\mu}_i = p_i^{\mu} + z n^{\mu} \nonumber \\
p^{\mu}_j &\rightarrow \hat{p}^{\mu}_j = p_j^{\mu} - z n^{\mu} \ ,
\label{eq:genBCFWshift}
\end{align}
which preserves momentum conservation. Furthermore, $n^\mu$ must be
such that these momenta stay on-shell. Therefore the vector $n$ must
obey
\begin{equation} \label{eq:constraintnmu}
(p^{\mu}_i n_{\mu}) = (p^{\mu}_j n_{\mu}) = (n^{\mu} n_{\mu}) = 0 \ .
\end{equation}
These equations do not have a solution for real $n_\mu$, but do for
complex momenta, as can be easily verified by going to the
center-of-mass frame. This shift changes the amplitude into a
function of $z$.

The amplitude we are interested in is given by $A_n(z=0)$, which can
be obtained by a contour integration around a contour which only
encompasses $z=0$,
\begin{equation}
A_n (0) = \oint_{z=0} \frac{A_n(z)}{z} d z \ .
\end{equation}
If the contour is now pulled to infinity one encounters various
poles at finite values of $z$ and a possible residue at infinity,
\begin{equation}\label{eq:pullingcontourtoinf}
A_n (0) = \oint_{z=0} \frac{A_n(z)}{z} dz = - \left\{
\sum \mathrm{Res}_{z=\textrm{finite}} +\sum \mathrm{Res}_{z= \infty} \right\} \ .
\end{equation}
In a unitary theory, the only poles possible at finite $z$ are
those at which physical particles become on-shell. Tree level
unitarity then guarantees that the residue at such a pole is the
product of two scattering amplitudes. In string theory at the disk
level this is (almost) manifest by the worldsheet `pinching'
argument illustrated in figure \ref{fig:factorizationprops}. Note
that the residues at the finite values of $z$ will involve
scattering amplitudes evaluated for \emph{shifted} values of the
momenta. Therefore, provided one can ignore the residue at infinity,
one arrives at recursion relations of the form
\begin{equation}
A_n(1,2,3 \ldots, n) = \sum_{r,h(r)} \sum_{k=2}^{n-2}
\frac{A_{k+1}(1,2,\ldots, \hat{i}, \ldots, k, \hat{P}_r)
A_{n-k+1}(\hat{P}_r, k+1, \ldots, \hat{j}, \ldots, n)}{\left(p_1 +
p_2 + \ldots + p_k \right)^2 +  m_r^2} \ . \label{eq:recursiongen}
\end{equation}
Here the label $r$ runs over all the states in the string theory,
$m_r$ is the corresponding mass and $h(r)$ symbolizes the sum over all possible
polarizations of the state. The momentum $\hat{P}_r$ for the `extra'
particle and its anti-particle in the amplitude is such that the
particle is on-shell. Note that in this expression one does not sum
over all kinematic channels: only those momentum invariants which
become $z$-dependent under the shift \eqref{eq:genBCFWshift} can
generate poles. Hence one does not obtain poles in the channels for
which the momentum invariant involves the sum of the shifted
momenta, since
\begin{equation}
 \hat{p}_i^{\mu} + \hat{p}_j^{\mu} = p_i^{\mu} + p_j^{\mu} \ .
\end{equation}
Therefore, while the resulting expression \eqref{eq:recursiongen}
incorporates some kinematic poles explicitly, it seems to lack
others. As will be shown below, this remark ties in very closely
with the old concept of `duality' in string theory.

In the following we will focus on proving in specific examples that
the residue at infinity is absent, thereby proving the recursion
relations. In passing it is noted that in the literature one often
refers to a `pole' at infinity spoiling recursion relations.
However, since there is a possibility that a function has a pole at
infinity but no residue, it is better to refer to this as a `residue
at infinity'.

\subsubsection*{General comments about BCFW recursion in string theory}
Since any (non-topological) string theory has an infinite tower of
massive states, there will in general be various infinite summations
throughout the calculation for poles at finite $z$ corresponding to
both the infinite set of massive states and the sums over the
polarization tensors of these massive states. This is a serious
drawback limiting the current calculational usefulness of the method
and we will make some comments about this further along in this
section. Another potential drawback of these recursion relations is that when
one starts with for instance a purely gluonic amplitude, then after
applying the recursion once one will need generically $2$ amplitudes
which involve one massive state each and a sum over its polarization
tensors. Therefore, the set of gluon amplitudes is generically not
closed under recursion, although the ordinary BCFW relations form a
subset of them and will be obtained in an $\al \rightarrow 0$ limit.
A comment related to this limit is that all the massive poles will
be at certain integer multiples of the string mass. Generically,
this will look like
\begin{equation}
(\mathcal{P} + \hat{p}_i)^2 = \frac{k}{\al} \ ,
\end{equation}
where $\mathcal{P}$ is a sum over momenta which does not involve
$\hat{p}_j$ and $k$ is a negative integer. The corresponding value of $z$
is
\begin{equation}
 z_{\textrm{pole}}  =  \frac{1}{2 (n \cdot \mathcal{P})} \left(\frac{k}{\al} - (\mathcal{P} + p_i)^2\right) \ .
\end{equation}
Hence, these poles will move to $z= \infty$ if $\al \rightarrow 0$,
leading generically to non-zero residues there. This is also to be
expected: as pointed out in \cite{ArkaniHamed:2008yf}, one can
interpret the shift \eqref{eq:genBCFWshift} for very large values of
$z$ as describing a hard particle moving through a soft background.
The $\al \rightarrow 0$ limit should more appropriately be described
without dimensions as the limit in which \emph{all} momentum invariants
are smaller than $1$. Hence first taking this limit and then
studying a `hard' particle for which some momentum invariants will
be larger than one will generically lead to inconsistencies: these
are precisely contained in the poles at infinity. Hence one should
always first derive recursion relations and then apply an expansion
in $\al$, if possible.  This is related to the fact that the $z
\rightarrow \infty$ limit is thought to be connected to the UV
properties of the theory under study: while string theory has
excellent behavior in the UV being (at least one (string)loop)
finite, any effective field theory derived from the string will
generically have bad behavior since it is non-renormalisable. Note
that this is also a hand-waving argument why the residue at infinity
should be expected to be absent for \emph{any} string theory
amplitude; that is, for any amplitude for any particle type in any
unitary string theory.

Another indirect argument for the correctness of
\eqref{eq:recursiongen} is that repeated application of this
relation (i.e. disregarding any potential residues at infinity)
allows one to express any string theory amplitude in terms of just
three-point on-shell amplitudes. This is reminiscent of open bosonic
string field theory, although making this precise would probably
require much work. On the other hand, note that closed bosonic
string field theory does not share this property. We will take both
the open string field theory argument and more importantly the UV behavior argument
as a prime indication that string theory amplitudes may obey
\eqref{eq:recursiongen}. As emphasized, the verification of this
statement depends on the absence of residues at infinity: we will
verify this explicitly in a number of examples.

\subsection{Veneziano amplitude}
Before proceeding to a more general argument and superstring gluon
amplitudes, it is instructive to note that even the mother of all
string theory amplitudes obeys recursion relations. This is of
course the Veneziano amplitude for four tachyons in open bosonic
string theory. The full Veneziano amplitude is simply given by
\begin{equation}\label{eq:fullVenezianoamp}
A_{4} = A_4^{\textrm{part}}(s,t) + A_4^{\textrm{part}}(t,u) +
A_4^{\textrm{part}}(u,s) \ ,
\end{equation}
where
\begin{equation}
A_4^{\textrm{part}}(s,t) = \frac{\Gamma(\al s -1) \Gamma(\al t
-1)}{\Gamma(\al (s+t) -2)} \ .
\end{equation}
Here $s = (p_1 + p_2)^2$, $t = (p_1 + p_3)^2$ and $u = (p_2 +
p_3)^2$ are the customary Mandelstam variables. This formula and its
$n$-particle generalizations have several remarkable properties
detailed in the literature, see e.g. \cite{DiVecchia:2007vd} for a
recent review. Here a shift of particles $1$ and $2$ is considered
for which
\begin{equation}\label{eq:explicitshiftsvenez}
\hat{s} = s \: , \quad \hat{t} = t - 2 z (p^{\mu}_3 n_{\mu})\: ,
\quad \hat{u} = u + 2 z (p^{\mu}_3 n_{\mu}) \ .
\end{equation}
By construction, the vector $n_{\mu}$ obeys equation
\eqref{eq:constraintnmu}. Note that every other shift is equivalent
by momentum conservation. Before continuing it is convenient to lose
the $2 (p^{\mu}_3 n_{\mu})$ factor altogether by rescaling the
integration variable $z$,
\begin{equation}
z' = 2 \al (p^{\mu}_3 n_{\mu}) z \ .
\end{equation}
Note that the fact that this can be done is a kinematical accident
for four particles.

There are two different types of term in \eqref{eq:fullVenezianoamp}
after the shift \eqref{eq:explicitshiftsvenez}: there is one term in
which two momentum invariants $(t,u)$ are shifted, and two terms in
which one momentum invariant is shifted. Let us first consider the
last type of contribution, say the first term in \eqref{eq:fullVenezianoamp}
for which \eqref{eq:pullingcontourtoinf} gives
\begin{equation}\label{eq:fourpointvenezrecur}
A_4^{\textrm{part}}(s,t) = - \textrm{Res}_{z' =
\infty}\left(\frac{A^{\textrm{part}}_{4}(s,t,z')}{z'}\right) +
\sum_{n=0}^{\infty} \frac{(-1)^n}{\Gamma(n+1)} \frac{\Gamma(\al s -
1)}{\Gamma(\al s -1 -n)} \frac{1}{\al t -1+ n} \ .
\end{equation}
As long as $\al s$ is not evaluated at one of the
resonances, the infinite sum on the right hand side of this equation
is equivalent to $A_4^{\textrm{part}}(s,t)$ by an identity for the
Beta function. This identity can for instance be derived by using
the integral representation $B(a,b) = \int_{0}^1 x^{a-1}
(1-x)^{b-1}$ and expanding $(1-x)^{b-1}$ using Newton's binomial
theorem. Note that this particular identity appears for instance in
\cite{Green:1987sp}, eq. (1.1.15), which identifies that equation as BCFW on-shell recursion avant la lettre. This proves explicitly in this particular case that
\begin{equation}
\textrm{Res}_{z' = \infty}\left(\frac{A^{\textrm{part}}_{4}(s,t,z')}{z'}\right) =0 \ .
\end{equation}
The same argument goes through for the third term in \eqref{eq:fullVenezianoamp},
$A_4^{\textrm{part}}(u,s)$.  Actually, the infinite summation in the sum in \eqref{eq:fourpointvenezrecur}
is usually used to illustrate the `duality' of an open string amplitude: the infinite series has only explicit poles in $t$, but when summed also exhibits poles in $s$. This is just the observation
that a string worldsheet can be pinched in multiple ways.

The contribution for which two momentum invariants are shifted,
$A_4^{\textrm{part}}(u,t)$, is more tricky as it reads,
\begin{align}
A_4^{\textrm{part}}(u,t) =  & -\textrm{Res}_{z' = \infty}\left(\frac{A^{\textrm{part}}_{4}(u,t,z')}{z'}\right)  \nonumber \\
 &  + \sum_{n=0}^{\infty} \frac{(-1)^n}{\Gamma(n+1)} \frac{\Gamma(\al (u+t) - 2+n)}{\Gamma(\al (u+t) -2)}  \left(\frac{1}{\al t -1+ n} + \frac{1}{\al u -1+ n} \right) \label{eq:trickyreccur} \ .
\end{align}
This particular sum equals $A_4^{\textrm{part}}(u,t)$ again as can be checked directly through a calculation in Maple. Hence
\begin{equation}\label{eq:eqwithprov}
\textrm{Res}_{z' = \infty} \left(\frac{A_{4}(u,t,z')}{z'}\right) = 0
 \ .
\end{equation}
Note that even without Maple, one can check that does have all the right poles. In addition, we have checked analytically up to order $\al^1$ and numerically up to order $\al^{10}$ that this is true in the sense that numerically the difference between the sum and the left hand side seems to converge to zero by taking into account more and more terms of the sum, leaving $u,t$ as free variables. Furthermore, in the case $u=-t$ it can be proven by hand since then on the left hand side of the equation Euler's reflection identity,
\begin{equation}\label{eq:eulerreflecid}
\Gamma(z) \Gamma(- z) = - \frac{\pi}{z \sin(\pi z)} \ ,
\end{equation}
can be used, while the right hand side can be explicitly molded into the Laurent series of this function. Note that a general order by order in $\al$ comparison involves very interesting identities between sums over (powers of) harmonic numbers and explicit zeta function values, also known as Euler sums. We are actually unsure whether all of these are known. Note that in a color ordered amplitude all three contributions in
\eqref{eq:fullVenezianoamp} would be independent, so it is encouraging that the residue at infinity terms vanishes separately for these three terms. However, one would like a more direct argument for the absence of residues at infinity so the above result could be derived. This will be constructed below.

\subsection{Absence of residues at infinity for four-point functions}
For the above shifts in the $(s,t)$ channel one can study the
integral at infinity directly\footnote{taking inspiration from \cite{bateman}, section 1.4.}. For this integral to be well-defined, let us
pull the contour in \eqref{eq:pullingcontourtoinf} to a large but
finite value of $z'$. For definiteness, let us specify the contour
as a circle of radius $R_k$, where $R_k$ is a half-integer
($R_k=k+\frac{1}{2}$, $k \in \mathbb{N}$), so we obtain
\begin{equation}
A^{\textrm{part}}(0)(s,t) = \sum_{n=0}^{k}
\frac{(-1)^n}{\Gamma(n+1)} \frac{\Gamma(\al s - 1)}{\Gamma(\al s -1
-n)} \frac{1}{\al t -1+ n} + \oint_{C_{R_k}} \frac{1}{z'}
\frac{\Gamma(\al s -1) \Gamma(\al t + z' -1)}{\Gamma(\al (s+t) +z'
-2)} dz   \ .
\end{equation}
For very large radius, we can approximate the integrand of the
contour integral by a well-known formula
\begin{equation}\label{eq:strilingapprox}
\frac{1}{z'} \frac{\Gamma(\al t + z' -1)}{\Gamma(\al (s+t) +z' -2)}
= (z')^{-\al s} \left(1 + \frac{1}{2 z'} (1-\al s) (2 \al t + \al
s-3) + \mathcal{O}\left(\frac{1}{(z')^2}\right) \right) \ .
\end{equation}
This formula is usually used to display Regge behavior of the string
theory amplitudes. It is actually only valid for $-\pi < \arg(z) <
\pi$, but the missing point may simply be deleted from the contour
integral as it is a measure zero set. Since the integrand is
certainly analytic in a neighborhood of this point, there is no
possibility of a contribution to the integral from this set. The resulting elementary integral can be calculated since generally,
\begin{equation}
\oint_{C_R} \frac{1}{z^a} dz  = i R^{1-a} \int d\theta e^{i (1-a)\theta}  =  R^{1-a} \frac{1}{1-a} \left(e^{i 2 \pi (1-a) } - 1 \right) \ .
\end{equation}
Hence, as long as ${\rm \Re} \left(1-a\right) < 0$ the elementary integral vanishes for large radius. In the case under consideration, this translates to
\begin{equation}\label{eq:condforrec}
{\rm \Re} \left(\al s\right) > 1 \ .
\end{equation}
Note that this condition neatly evades all the resonant poles in the amplitude. This is also needed as the sum does not contain explicit poles in the $s$ channel.

For the shift in the $(u,t)$ contribution one obtains similarly the integral
\begin{equation}
\textrm{Res}_{z' =
\infty}\left(\frac{A^{\textrm{part}}_{4}(u,t,z')}{z'} \right) =
\oint_{C_R} \frac{1}{z'} \frac{\Gamma(\al u - z'  -1) \Gamma(\al t +
z' -1)}{\Gamma(\al (u+t) -2)} dz \ .
\end{equation}
Using \eqref{eq:eulerreflecid}, this can be written as
\begin{align}
\textrm{Res}_{z' = \infty}\left(\frac{A^{\textrm{part}}_{4}(u,t,z')}{z'} \right) =
\oint_{C_R} \frac{1}{z'} & \frac{1}{\Gamma(\al (u+t) -2)} \frac{\Gamma(\al t + z' -1)}{\Gamma( 1 - \al u + z')} \nonumber \\
& \frac{\pi}{(\al u - z'  -1) \sin\left(\pi (\al u - z'  -1)\right)} dz \ .
\end{align}
For large $z'$, we can apply \eqref{eq:strilingapprox} again.
However, the secant function also has a non-trivial large $z'$
expansion as it vanishes exponentially along any direction not along
the real axis, and is analytic along that axis. Hence the behavior
at infinity is actually far better than for the other case: the
integrand vanishes exponentially on the contour up to a measure zero
set (supported on two points in this case), in the neighborhood of
which the integrand is analytic. This reflects itself in the fact
that in this case the resulting recursion relation, \eqref{eq:trickyreccur}, contains the two possible series of poles
explicitly so there is no need for a restriction like \eqref{eq:condforrec}. Hence \eqref{eq:eqwithprov} is actually
true without any provisos.

These results have a very important immediate extension: if the Beta
function had been multiplied by \emph{any} rational function of $z'$
or more generally, any function which has a Laurent series at
infinity, the same result would follow. The only thing which changes
is that if the prefactor diverges, $\sim z^k$ for some positive $k$,
then the constraint \eqref{eq:condforrec} on $\al s$ would
change slightly. However, as long as the appropriate reality condition is satisfied, there
are no problems. Hence we expect that the above derivation
holds for \emph{any} four particle amplitude in \emph{any} open
string theory with arbitrary external states. In particular, we
expect recursion to work for the gluon amplitude in the superstring,
as we will verify below. Some subtleties related to the $\al$ expansion will also be discussed there.

\subsubsection*{Closed strings: Virasoro-Shapiro amplitude}
Although this article mainly deals with open strings, it is hard to
resist generalizing the above derivation of the absence of a residue
at infinity to the scattering of four closed string tachyons in
bosonic string theory. This particular amplitude is known as the
Virasoro-Shapiro amplitude and is given by,
\begin{equation}\label{eq:virasoroshapiro}
A_4(s,t,u) = \frac{\Gamma\left(\frac{\al s - 1}{2}\right) \Gamma\left(\frac{\al t -
1}{2}\right) \Gamma\left(\frac{\al u - 1}{2}\right) }{\Gamma\left(1- \frac{(\al s
-1)}{2} \right) \Gamma\left(1- \frac{(\al t -1)}{2} \right)
\Gamma\left(1- \frac{(\al u -1)}{2} \right)} \ .
\end{equation}
The previous analysis of the shift in the $(u,t)$ channel for the
open string amplitude, resulting in \eqref{eq:trickyreccur},
can be adapted for the above amplitude using the same shift. One then finds
\begin{align}
\label{eq:A4uts}
A_4(s,t,u) & =   -\textrm{Res}_{z' = \infty}\left(\frac{A_{4}(s,t,u,z')}{z'}\right)  \nonumber \\
   + \sum_{n=0}^{\infty} &  \frac{(-1)^n}{\Gamma(n+1)}
 \frac{\Gamma \left(\frac{\al s - 1}{2}\right)}{\Gamma \left(1- \frac{(\al s-1)}{2} \right)}
\frac{\Gamma\left(\frac{\al (t+u) - 2+n}{2}\right)}{\Gamma \left(2- \frac{[\al (t+u) +n ]}{2} \right)}
 \left(\frac{1}{\al t -1+ n} + \frac{1}{\al u -1+ n} \right)  \ .
\end{align}
We have not been able to analytically evaluate the sum in this expression, but we have checked numerically that the difference between the sum and the left hand side seems to converge to zero by taking into account more and more terms of the sum. One can also show the absence of the residue at infinity more directly, following the same argument as above by considering
\begin{align}
\textrm{Res}_{z' = \infty}\left(\frac{A_{4}(s,t,u,z')}{z'}\right) & = \frac{\Gamma\left(\frac{\al s - 1}{2}\right)}{\Gamma\left( \frac{3 -\al s }{2} \right) } \textrm{Res}_{z' = \infty} \left(\frac{1}{z'} \frac{ \Gamma\left(\frac{\al t -
1}{2}\right) \Gamma\left(\frac{\al u - 1}{2}\right) }{\Gamma\left(\frac{3 - \al t - z'}{2} \right) \Gamma\left(\frac{3 - \al u + z'}{2} \right)} \right) \nonumber  \\
& = \frac{\Gamma(\al s - 1)}{\Gamma\left( \frac{3 -\al s }{2}
\right) } \textrm{Res}_{z' = \infty} \left( \frac{1}{z'} \left(\frac{z'}{2}\right)^{-\frac{\al s}{2}}\left(-\frac{z'}{2}\right)^{-\frac{\al s}{2}}
\left(1 + \mathcal{O}(\frac{1}{z'}) \right) \right) \ .
\end{align}
For an appropriate reality condition on $s$, the integrand vanishes at infinity as above. In addition, the integrand is
a well-defined analytic function in the neighborhood of the measure zero set on which the asymptotic series is not valid. Hence the
residue integral at infinity vanishes for this particular amplitude
and the Virasoro-Shapiro amplitude obeys on-shell recursion
relations. Although we will not write them out explicitly here, note
that the infinite sum one obtains indeed incorporates all the poles
of the full amplitude explicitly. By the same extension as before,
any rational function of $z'$ times \eqref{eq:virasoroshapiro} is
therefore also expected to obey recursion relations since the
residue at infinity is absent there. This argument therefore
applies to all closed string four-point amplitudes, as they have this
generic shape.

\subsubsection*{Extension to higher point functions}

Higher point functions are more complicated, but the general argument given above should continue to hold. The reason this is the case is that the generic integral appearing for, say the $n$-point function can always be decomposed into infinite sums over Gamma functions. Assuming the argument given above continues to commute
with the infinite sums, it is very hard to see what could go wrong. Of course, it would be very interesting to make this more precise, especially as we will see that the recursion relation even for the $5$-gluon amplitude is fairly non-trivial.

\subsection{The four- and five-point gluon amplitudes}
As detailed earlier in equation \eqref{eq:fourpoints}, the $4$-point
gluon amplitude in four dimensions is given by
\begin{equation}
A_4(1^-,2^-,3^+,4^+) = \frac{\langle 12 \rangle^4}{\langle 12
\rangle \langle 23 \rangle \langle 34 \rangle \langle 41 \rangle}
\frac{\Gamma(1+\al s)\Gamma(1+\al t)}{\Gamma(1+\al s + \al t)} \ .
\label{eq:4gluonMHV}
\end{equation}
We can now follow the same steps as above using again the shifts
in eq.~\eqref{eq:explicitshiftsvenez}.

\subsubsection*{A subtlety}
At this point one should be slightly more careful about the
helicities of the states in the gluon amplitude when applying the
shift. In four dimensions, there are two solutions to the
constraints \eqref{eq:constraintnmu} for the vector $n^{\mu}$:
\begin{equation}\label{eq:choiceofn}
n^{\mu} = p_1^{\alpha} p_2^{\dalpha} \quad \textrm{or} \quad n^{\mu}
= p_2^{\alpha} p_1^{\dalpha} \ .
\end{equation}
We will pick the first one, so that the spinors transform as
\begin{align}
\label{eq:shift}
p_1^{\dalpha} & \rightarrow  p_1^{\dalpha} - z p_2^{\dalpha} \nonumber \\
p_2^{\alpha} & \rightarrow  p_2^{\alpha} + z p_1^{\alpha} \ .
\end{align}
This leads to the transformations
\begin{equation}
s \longrightarrow s \: , \quad t \longrightarrow t - z \langle 13
\rangle [23] \quad \mathrm{and} \quad \langle 23 \rangle
\longrightarrow \langle 23 \rangle + z \langle 13 \rangle \ .
\label{eq:kin-var-shifts}
\end{equation}
Now one finds by explicit inspection of the Yang-Mills factor in
\eqref{eq:4gluonMHV} that it vanishes for $z \rightarrow \infty$ if
and only if the helicities of $1$ and $2$ are $(+,+)$, $(-,-)$ or
$(+,-)$ respectively. The shift in helicities $(-,+)$ is
`well-behaved' for the other solution in \eqref{eq:choiceofn}.
However, as we will see below, in string theory the situation is
better: also the bad shift actually works.

\subsubsection*{Recursion}

Below we focus on the $A_4(--++)$ amplitude; the other cases are
similar, keeping the above subtlety in mind. Applying the
transformations \eqref{eq:kin-var-shifts} to \eqref{eq:4gluonMHV}
one finds
\begin{equation}
A_4(z) = \frac{\langle 12 \rangle^4}{\langle 12 \rangle \big(
\langle 23 \rangle + z \langle 13 \rangle \big) \langle 34 \rangle
\langle 41 \rangle} \frac{\Gamma(1+\al s) \Gamma(1+\al t - \al z
\langle 13 \rangle [23])}{\Gamma(1+\al s + \al t - \al z \langle 13
\rangle [23])} \ .
\end{equation}
As a function of $z$, $\frac{A_4(z)}{z}$ has the following poles,
apart from the one at $z=0$,
\begin{itemize}
\item at $z = -\frac{\langle 23 \rangle}{\langle 13 \rangle}$,
with residue
\begin{equation}
\mathrm{Res} \left( \textstyle{\frac{A_4(z)}{z}}, z =
-\textstyle{\frac{\langle 23 \rangle}{\langle 13 \rangle}} \right) =
- \frac{\langle 12 \rangle^4}{\langle 12 \rangle \langle 23 \rangle
\langle 34 \rangle \langle 41 \rangle} \ .
\label{eq:A4-gluonexchange}
\end{equation}
This pole corresponds to the exchange of a gluon in the pinched disk
diagram picture, and the residue yields exactly the Yang-Mills
result. The latter fact is entirely expected since the residue must
factor into two 3-gluon amplitudes; however, as these amplitudes
receive no $\al$-corrections in superstring theory (cf.
\eqref{eq:threepoints}), their product must reproduce the Yang-Mills
result.

\item at $z = \frac{k + \al t}{\al \langle 13 \rangle [23]}$, $k
\in \mathbb{N}$, with residues {\small \begin{equation} \mathrm{Res}
\left( \textstyle{\frac{A_4(z)}{z}}, z = \textstyle{\frac{k + \al
t}{\al \langle 13 \rangle [23]}} \right) = \frac{(-1)^k}{k!}
\frac{\langle 12 \rangle^4}{\langle 12 \rangle \langle 23 \rangle
\langle 34 \rangle \langle 41 \rangle} \frac{(\al s)(\al s -1)
\cdots (\al s - k+1)}{t + k/\al} t \ .
\end{equation}}
\end{itemize}
Summing the residues yields
\begin{eqnarray}
&\phantom{=}& \hspace{-3cm} \mathrm{Res} \left(
\textstyle{\frac{A_4(z)}{z}}, z = -\textstyle{\frac{\langle 23
\rangle}{\langle 13 \rangle}} \right) + \sum_{k=1}^\infty
\mathrm{Res} \left( \textstyle{\frac{A_4(z)}{z}}, z =
\textstyle{\frac{k + \al
t}{\al \langle 13 \rangle [23]}} \right) \nonumber \\
&=& - \frac{\langle 12 \rangle^4}{\langle 12 \rangle \langle 23
\rangle \langle 34 \rangle \langle 41 \rangle} \frac{\Gamma(1+\al
s)\Gamma(1+\al t)}{\Gamma(1+\al s + \al t)} \ .
\end{eqnarray}
Using \eqref{eq:4gluonMHV} and \eqref{eq:pullingcontourtoinf} this shows in particular
\begin{equation}
\oint_\infty \frac{A_4(z)}{z} \, dz = 0 \ .
\end{equation}

\subsubsection*{The `bad' shift}
Let us now consider the `bad' shift for particles $2$ and $3$
(i.e. $p_2^{\dalpha} \rightarrow  p_2^{\dalpha} - z p_3^{\dalpha}$
$p_3^{\alpha}  \rightarrow  p_3^{\alpha} + z p_2^{\alpha} $) applied
the same amplitude $A_4(++--)$. For this shift we obtain
\begin{equation}
A_4(z) = (\braket{12} + z \braket{13})^2 (\sbraket{34} - z
\sbraket{24})^2 \frac{\Gamma(\al s + z \sbraket{12} \braket{13})
\Gamma(\al t )}{\Gamma(1+ \al t + \al s + z \sbraket{12} \braket{13})} \ .
\end{equation}
In the same way as before the residue of the
function $\frac{A_4(z)}{z}$ can be considered. In order to prove that the residue at
infinity is absent directly, as we expect on the basis of our
general argument, we should have
\begin{equation}
A_{4}(0) = - \sum_{n=1}^{\infty} \frac{(-1)^n}{\al^2 \Gamma(n+1)}
\frac{n^2\left(n \sbraket{2 3} + \al \sbraket{12}\left(\braket{12}
\sbraket{24} + \braket{1 3} \sbraket{3 4} \right)
\right)^2}{\braket{1 3}^2 \sbraket{1 2}^4} \frac{1}{n + s}
\frac{\Gamma(\al t)}{\Gamma(1 + \al t -n)} \ .
\end{equation}
That this is actually true can be verified explicitly by performing
the sum using Mathematica. Note that an $\al$ expansion of the separate terms
in the infinite sum is ill-defined as it involves the sum
\begin{equation}
\sim \sum_{n=1}^{\infty} \frac{n^2}{\al^2}  +
\mathcal{O}(\frac{1}{\al}) \ ,
\end{equation}
which is a well-known diverging sum in string theory.  This is related to the fact that according to the analysis above (specifically eq.~\eqref{eq:condforrec}), recursion should only work when
\begin{equation}\label{eq:badshiftbadbehavior}
{\rm Re}\left(3-\al t \right) <0 \ .
\end{equation}
Obviously, this condition cannot be satisfied when $\al =0$.

\subsubsection*{Expansion in $\al$ within the recursion relation}
For the 'good' shift the equivalent condition \eqref{eq:condforrec} is met easily if $\al=0$ as the condition reads,
\begin{equation}\label{eq:goodshiftgoodbehavior}
{\rm Re}\left(-1-\al t \right) < 0 \ .
\end{equation}
Noting that
\begin{equation}
s t \frac{\langle 12 \rangle^4}{\langle 12 \rangle \langle 23
\rangle \langle 34 \rangle \langle 41 \rangle} = - \langle 12
\rangle^2 [34]^2 \ ,
\end{equation}
one furthermore finds from \eqref{eq:4gluonMHV} the form
\begin{eqnarray}
&\phantom{=}& \hspace{-1.5cm} A_4(1^-,2^-,3^+,4^+) \nonumber \\
&=& \frac{\langle 12 \rangle^4}{\langle 12 \rangle \langle 23
\rangle \langle 34 \rangle \langle 41 \rangle} + \sum_{k=1}^\infty
\frac{(-1)^{k-1}}{k!} \frac{\al (\al s -1)\cdots (\al s -k+1)}{t +
k/\al} \langle 12 \rangle^2 [34]^2
\label{eq:decomp-A4(1)} \\
&=& \frac{\langle 12 \rangle^4}{\langle 12 \rangle \langle 23
\rangle \langle 34 \rangle \langle 41 \rangle} + \sum_{k=1}^\infty
\frac{1}{k} \frac{\al}{t + k/\al} \langle 12 \rangle^2 [34]^2 +
\cdots \label{eq:decomp-A4(2)} \\
&=& \frac{\langle 12 \rangle^4}{\langle 12 \rangle \langle 23
\rangle \langle 34 \rangle \langle 41 \rangle} +
\frac{\pi^2}{6}\cdot \al^2 \langle 12 \rangle^2 [34]^2 +
\mathcal{O}(\al^3)  \ ,\label{eq:decomp-A4(3)}
\end{eqnarray}
where the `$\cdots$' in \eqref{eq:decomp-A4(2)} refer to an
$\al$ expansion of the numerator in the sum over $k$ in
\eqref{eq:decomp-A4(1)}. We note from \eqref{eq:decomp-A4(2)} that
at each mass level $k$ the $\mathcal{O}(\alpha'^2)$ correction takes
the universal form $\frac{\alpha'^2 \langle 12 \rangle^2 [34]^2}{k^2}$.
Furthermore, this correction can be factorized into two `effective
vertices' $\frac{\alpha'(p_{1\alpha} p_{4\dalpha})
(p_{1\beta} p_{4\dbeta})}{k}$ and
$\frac{\alpha'(p_{2\alpha} p_{3\dalpha})
(p_{2\beta} p_{3\dbeta})}{k}$, each corresponding to a
coupling between two gluons and a state with mass squared
$k/\alpha'$. This suggests in principle that one can compute
higher-order corrections in $\alpha'$ by use of further effective vertices.
However, in practice this is not feasible, partly because these
vertices become increasingly complicated, and partly because at a
given order in $\alpha'$ one needs to sum over several such
vertices, cf. eq. \eqref{eq:decomp-A4(1)}.

The example of the four-point gluon amplitude also illustrates a general remark we would like to make about $\al$ expansion versus recursion relations. As already noted above, on physical grounds one expects that one should be careful about `commuting' them. Above we explicitly see how this expectation works in a particular example. Based on this, we strongly suspect that for massless particles deriving recursion relations and applying $\al$ expansion commute only for what in field theory are `well-behaved' shifts (see the discussion below eq. \eqref{eq:kin-var-shifts}). Specifically, we expect that when one derives recursion relations for amplitudes involving only massless particles, only a constraint on the kinematic variables of the form   \eqref{eq:goodshiftgoodbehavior} will be encountered for a BCFW shift that works in field theory and \eqref{eq:badshiftbadbehavior} for a shift that does not work in field theory.

\subsection*{The 5-gluon amplitude}
\label{sec:5lguonamp}

We now apply the same procedure to the more complicated case of the 5-gluon
MHV amplitude which is given by
\begin{equation}
\label{eq:5gluon}
A_5(1^-,2^-,3^+,4^+,5^+) = \frac{\langle 12 \rangle^4}{\langle 12
\rangle \langle 23 \rangle \langle 34 \rangle \langle 45 \rangle
\langle 51 \rangle} \big( \alpha'^2 s_{51} s_{23} f_1 + \alpha'^2
[12] \langle 23 \rangle [35] \langle 51 \rangle f_2 \big) \ ,
\end{equation}
where {\footnotesize \begin{equation} f_1 = \frac{\Gamma(\alpha'
s_{23}) \Gamma(\alpha' s_{51}) \Gamma(\alpha' s_{34} + 1)
\Gamma(\alpha' s_{45}+1)}{\Gamma(\alpha' s_{23} + \alpha' s_{34} +
1) \Gamma(\alpha' s_{51} + \alpha' s_{45} + 1)} \phantom{.}_3 F_2
\left[ \begin{array}{c} \alpha' s_{23}, \alpha' s_{51}, -\alpha' s_{35} \\
\alpha' s_{23} + \alpha' s_{34} + 1, \alpha' s_{51} + \alpha' s_{45}
+ 1
\end{array} ; 1\right] \ ,
\end{equation}}
and {\footnotesize \begin{equation} f_2 = \frac{\Gamma(\alpha'
s_{23}+1) \Gamma(\alpha' s_{51}+1) \Gamma(\alpha' s_{34} + 1)
\Gamma(\alpha' s_{45}+1)}{\Gamma(\alpha' s_{23} + \alpha' s_{34} +
2) \Gamma(\alpha' s_{51} + \alpha' s_{45} + 2)} \phantom{.}_3 F_2
\left[ \begin{array}{c} \alpha' s_{23} + 1, \alpha' s_{51} +1,
-\alpha' s_{35}+1 \\ \alpha' s_{23} + \alpha' s_{34} + 2, \alpha'
s_{51} + \alpha' s_{45} + 2 \end{array} ; 1\right] \ .
\end{equation}}
In these equations $s_{ij} = (p_i + p_j)^2$ and the hypergeometric function
$\phantom{.}_3 F_2$ is defined through its series expansion
\begin{equation}
\label{eq:3F2}
{}_3 F_2 (a,b,c;d,e;z) = \sum_{k=0}^\infty \frac{ (a)_k (b)_k (c)_k}{(d)_k (e)_k}
\frac{z^k}{k!} \quad , \quad (x)_k \equiv \frac{\Gamma(x+k)}{\Gamma(k)} \ .
\end{equation}
We now apply the BCFW shift
\begin{equation}
\label{eq:bcfw5gl}
\left\{\begin{array}{lcl} p_{1 \alpha} &\longrightarrow& p_{1 \alpha} \\
p_{1 \dalpha} &\longrightarrow& p_{1 \dalpha} + z p_{2 \dalpha} \\
p_{2 \alpha} &\longrightarrow& p_{2 \alpha} - z p_{1 \alpha} \\
p_{2 \dalpha} &\longrightarrow& p_{2 \dalpha} \ ,
\end{array}\right.
\end{equation}
to the expression above (see appendix \ref{app:5pointshift} for
further details). For this particular shift, it follows that the
$z$-dependent parts of both $f_1$ and $f_2$ take the form
\begin{eqnarray}
f_{1,2}(z) & \sim  &  \frac{\Gamma(a+\kappa_1 z) \Gamma (b+\kappa_2 z)}{\Gamma(d+\kappa_1 z) \Gamma(e+\kappa_2 z)} \; {}_3 F_2 (a+\kappa_1 z,b + \kappa_2 z,c;d+ \kappa_1 z,
e + \kappa_2 z;1) \\ \label{eq:3F2b}
& & = \sum_{k=0}^\infty    \frac{\Gamma(a+k+\kappa_1 z) \Gamma (b+k+\kappa_2 z)}{\Gamma(d+k+\kappa_1 z) \Gamma(e+k+\kappa_2 z)} \frac{(c)_k}{k!} \label{eq:zbehavioroffi} \, .
\end{eqnarray}
From the expression for the analytically continued 5-gluon amplitude \eqref{eq:shiftedA5} it is clear that in
order to show $\oint_\infty \frac{A_5(z)}{z} \, dz = 0$ it is enough to show that
$\oint_\infty \frac{f_i(z)}{z} \, dz = 0$ and $\oint_\infty f_i(z) \, dz = 0$ for $i=1,2$. The two cases are treated
completely analogously, and we focus here on the latter one.

Thus we aim to show $\oint_\infty f_i(z) \, dz = 0$. From \eqref{eq:zbehavioroffi} it suffices to show\footnote{Note that in the following we assume the validity of $\sum_k \oint = \oint \sum_k$.}
\begin{equation}
\oint_\infty \frac{\Gamma(a+\kappa_1 z) \Gamma (b+\kappa_2 z)}{\Gamma(d+\kappa_1 z) \Gamma(e+\kappa_2 z)} \, dz = 0 \: .
\label{eq:vanishingintegral}
\end{equation}
Redefining $z' = \kappa_1 z$ and putting $\kappa\equiv\kappa_2/\kappa_1$,
from Stirling's formula the integrand has the large $z'$ behavior
\begin{equation}
\label{eq:3F2largez}
\frac{\Gamma(a+ z') \Gamma (b+\kappa  z')}{\Gamma(d+ z') \Gamma(e+\kappa z')} \simeq (z')^{a-d} (\kappa z')^{b-e} \left[ 1 + {\cal{O}}(z'{}^{-1}) \right] \ ,
\end{equation}
where the corrections are a Laurent series in $z'$ and one needs to require
$ -\pi < \arg(z') < \pi$ and $ -\pi < \arg (\kappa z') < \pi $. (In further detail, both for $f_1$ and $f_2$ one finds $a-d= -\al s_{34} -1 $ and $b-e= -\al s_{45}-1 $, while
$\kappa = -\langle 51 \rangle [ 25]/( \langle 13\rangle [32]) $.) Assuming $\arg(\kappa) \neq \pi$ we may now compute the contour integral of the left hand side of \eqref{eq:3F2largez} around infinity in the
following way. Considering first the contour to be a circle of finite radius, the points corresponding
to the angles $-\pi$ and $\pi - \arg(\kappa)$ divide the circle into two arcs. Without loss of generality take the angular range of the larger arc to be from $\pi - \arg(\kappa)$ to $-\pi$ (i.e. so that the arc is contained mainly
in the upper half plane). The integral over this arc (with the endpoints excluded) in the large radius limit may then be computed using the Stirling approximation \eqref{eq:3F2largez}. To compute the integral over the remaining part of the circle (including the points $-\pi$ and $\pi - \arg(\kappa)$) we apply Euler's reflection identity to the integrand
which then becomes
\begin{equation}
 \frac{\Gamma(1-d - z') \Gamma(1-e-\kappa z')}{\Gamma(1-a -z') \Gamma (1-b- \kappa z')} \frac{\sin[ \pi(d+z')]}{\sin [ \pi (a+z')] } \frac{ \sin[ \pi(e + \kappa z') ]}{\sin [\pi (b + \kappa z')]} \: .
\end{equation}
The asymptotic behavior of this function is $(-z')^{a-d} (-\kappa z')^{b-e}$ when $-\pi < \arg(-z') < \pi$
and $-\pi < \arg(-\kappa z') < \pi$. One may choose the full contour in \eqref{eq:vanishingintegral} as an arbitrarily large circle avoiding the zeroes of $\sin[\pi(a+z')]$ and $\sin[\pi(b+\kappa z')]$. On the arc of this circle running from
$-\pi$ to $\pi - \arg(\kappa)$ the $\frac{\sin[ \pi(d+z')]}{\sin [ \pi (a+z')]} \frac{ \sin[
\pi(e + \kappa z') ]}{\sin [\pi (b + \kappa z')]}$ factor is bounded, and the integral over the left hand side in
\eqref{eq:3F2largez} is thus numerically bounded by the integral $\oint |z'|^{a-d}|\kappa z'|^{b-e} \, dz'$. If
$ \rm{Re}\left(a + b -d-e +1 \right) <0 $ holds, then the integrand tends to zero as the radius is taken to infinity. In the case under study this amounts to
\begin{equation}
\rm{Re}\left(-1 - \al (s_{34} + s_{45})\right) <0 \ .
\end{equation}
This concludes the proof of \eqref{eq:vanishingintegral}. Moreover, as observed above for the four-point gluon amplitude,
the $\al \rightarrow 0$ limit commutes nicely with the constraint just derived.

As a further illustration of the result above, we consider in appendix \ref{app:5pointshift}
the poles of $A_5(z)$ and their corresponding residues explicitly. A general analysis is quite involved so that we focus on the residues to next-to-leading order in $\al$. We explicitly show how
to order $\al^2$ the residues of the finite poles reproduce the five-point
amplitude, implying that the residue at infinity is vanishing up to $\mathcal{O}(\alpha'^3)$.
The computation above thus proves that this holds exactly.

It is noteworthy to realize that every 5-point string theory amplitude involves on general grounds
hypergeometric functions ${}_3 F_2$. The crucial ingredient in the argument above is
that the BCFW shift acts as in \eqref{eq:3F2largez} on this function. Thus if given
a particular 5-point string amplitude one is able to finds a BCFW shift that has this property
then a BCFW recursion relation can be found. It would be interesting to study this
further and consider higher point functions.

\subsection{An attempt at truncating the BCFW recursion relations}

Since by the above argument the residue at infinity vanishes, the 5-gluon amplitude will equal (minus) the
sum over all the residues of $\frac{A_5(z)}{z}$. To make the recursion explicit, one would like to factorize the resulting expressions into amplitudes and a sum over polarizations. However, although unitarity in principle
guarantees the existence of such a factorization, it is neither
clear from the explicit forms of the residues how to extract the
factors, nor whether the expressions for such factors would yield
useful building blocks for recursion relations.

\begin{figure}[!h]
\begin{center}
\includegraphics[angle=0, width=0.8\textwidth]{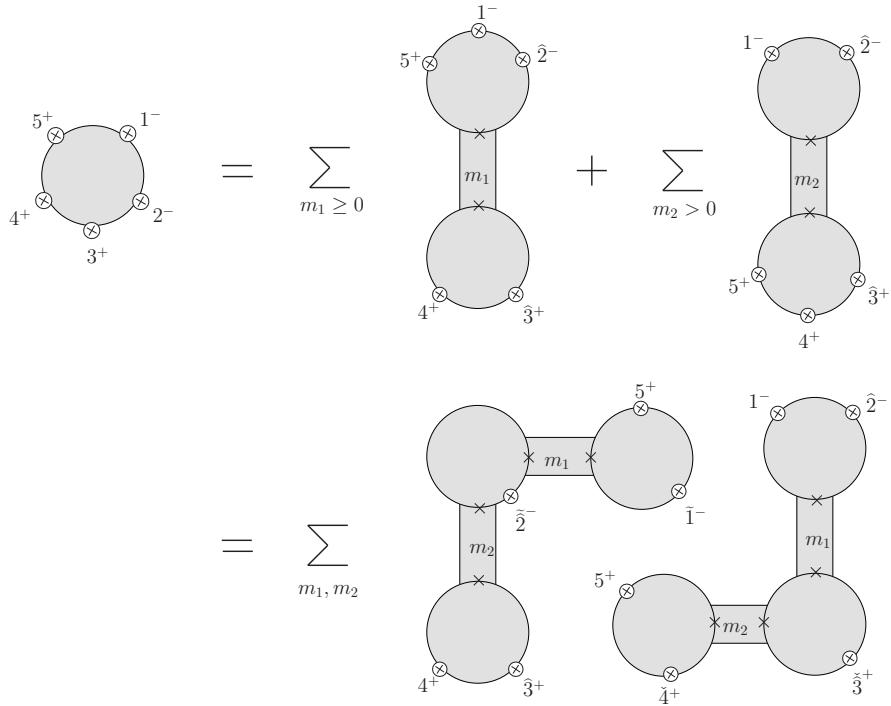}
\caption{A graphical illustration of $5$-gluon amplitude recursion
diagrams}\label{fig:recursion}
\end{center}
\end{figure}

It would be very interesting to obtain recursion relations order by order in $\al$. Inspired by the discussion following \eqref{eq:decomp-A4(1)}-\eqref{eq:decomp-A4(3)} that the full 4-gluon string amplitude is obtained by summing effective vertices
over massive states of the string, we consider first the following schematic recursion relation for the 4-gluon amplitude
\begin{equation}
A_4 (1^{h_1},2^{h_2},3^{h_3},4^{h_4}) = \sum_{r,h(r) }
A_3(1^{h_1},2^{h_2},P_r) \frac{1}{(k_1+k_2)^2 + m_r^2}
A_3(P_r,3^{h_3},4^{h_4}) \ , \label{eq:recursion-on-A4}
\end{equation}
where we have used the same notation as in \eqref{eq:recursiongen} for the sum over
string theory states.
We will now attempt to construct a truncation of the BCFW recursion relation for the $5$-gluon MHV amplitude $A_5(1^-,2^-,3^+,4^+,5^+)$ that would capture the $\al$ correction up to a certain order $k_0$. In agreement with the discussion in section \ref{sec:5lguonamp}, we draw the recursion relations schematically as in \ref{fig:recursion}. Note that in the first line we have to include an additional diagram compared to the usual Yang-Mills case (the rightmost diagram in the first line of figure \ref{fig:recursion}). Of course, this diagram vanishes for $m_2=0$. We can apply recursion again using the appropriate
generalization of \eqref{eq:recursion-on-A4} to arrive at the last line of figure \ref{fig:recursion}.

For the first graph in the bottom line, the strategy is now to split
the sum over $m_1,m_2$ into four contributions $m_1 = m_2 = 0$; $m_1
> 0, m_2 = 0$; $m_1 = 0, m_2 > 0$ and $m_1 > 0, m_2 > 0$. The first
contribution is just the Yang-Mills contribution; the sum over $m_1$
in the second contribution and the sum over $m_2$ in the third
contribution can be performed explicitly (equation
\eqref{eq:recursion-on-A4}), yielding again 4-gluon amplitudes. Note that up to now we have only used known amplitudes involving gluons.   Assuming for the moment that the fourth contribution (which involves amplitudes we have not encountered yet) starts appearing at an order in $\alpha'$ larger than $k_0$, one can
tentatively add the first three contributions and hope that this might reproduce the 5-gluon amplitude up to this order, thus providing a useful truncation of the recursion relations. Unfortunately, one finds discrepancies already at order $\alpha'^2$ where only part of the $\alpha'^2$ corrections
\eqref{eq:degfouranswerspin} are reproduced, but interestingly no spurious poles.

The obvious origin of the discrepancy is the contribution from the sum over $m_1 > 0, m_2 > 0$. However, that simply indicates that our truncation is too naive and new input is needed. There is a perhaps useful analogy here to one of the remarks at the end of section 3: soft and collinear limits only go some way to determine an amplitude as actual amplitudes are needed to fix some of the coefficients. For the $\al^2$ correction for instance, the $5$-point amplitude is needed and once this is known, the amplitude is fixed completely to this particular order in $\al$ by the kinematic limits. Motivated by this observation, there is a suspicion that a similar truncation to the one studied in this subsection actually does work from $6$-point amplitudes onwards (to order $\al^2$). This would be interesting to check.

\section{Conclusions and outlook}
In this article we have initiated a study of different approaches to
calculating amplitudes in string theory motivated by recent
developments in field theory. Using surprisingly basic
considerations, three results have been established regarding string
theory scattering, which will be commented on in turn. We
subsequently turn to more general comments.

\subsection{MHV, CSW and BCFW:}

\subsubsection*{MHV}

In section \ref{sec:alphaprime3}, we showed how the symmetry
properties of the $\al$ corrected MHV amplitudes, and in particular
their behavior under soft and collinear limits, severely restrict
the unknown polynomials $Q_m^{(N)}$ that appear in the correction
terms. In particular, we used a computer approach to find the
general $\al^3$ correction explicitly for any number $N$ of external
particles. It would of course be interesting to generalize this
approach to higher orders. As commented upon earlier in this paper, for $\al^4$ this might still be possible
using a more efficient computer approach along the lines of this paper (even though as we explained in
the text, additional exact input from $7$- and perhaps $8$-point amplitudes is also required), but for higher orders one would need
to further develop the analytic approach. These comments are further fleshed out at the end of chapter 3.

One comment which deserves some emphasis is that it would be useful to develop an idea of what the resulting polynomials would look
like. To achieve this, the spinor notation is probably the most convenient. For example, using this notation, $Q_2^{(N)}$ can be
written in the very simple form (\ref{eq:degfouranswerspin}). It would be nice if one could find a similarly simple form for the
polynomial $Q_3^{(N)}$ that we found in this paper, and use this to conjecture the general form of $Q_m^{(N)}$.

To round off this section, it would be extremely interesting to know if the full open superstring disk amplitude has localization properties on some natural object on twistor space or an extension thereof. Perhaps \cite{Berkovits:2008ic} can yield some insight here.

\subsubsection*{CSW}

In section \ref{sec:action}, scattering amplitudes have been derived
for the Dirac-Born-Infeld action. Despite the fact that the action
contains an infinite amount of vertices, it was shown that these
conspire to produce very simple results. This needs to be understood
more clearly, as the Rosly-Selivanov symmetry argument in \cite{Rosly:2002jt}
seems to suggest that the complete string theory should only
generate helicity conserving amplitudes. This constraint might point
the way to an all orders in $\al$ effective space-time action.

Concretely, it would be very interesting to see what happens to the $6$-point
MHV amplitude found by Stieberger and Taylor in \cite{Stieberger:2006bh} in the Abelian limit. The Rosly-Selivanov
argument suggests that this amplitude, properly summed over all
$120$ contributions, actually might vanish completely if the disk level
string theory amplitude indeed has a $U(1)$ duality symmetry. In
turn, this would imply very strong constraints on the derivative
corrections to the DBI action. Another potentially fruitful road to pursue would
be to generalize the scalar action we have found in a way that naturally incorporates both Veneziano amplitudes as
well as the DBI amplitudes. Perhaps here input from string field theory would be useful.

The non-Abelian effective action remains less tractable,
although the present article has yielded some more clues. Unfortunately it is still difficult to connect all the dots. Deriving
scattering amplitudes from the known actions would still be
interesting to check these actions against string theory amplitudes.
Also the question of non-Abelian string theory CSW rules remains
open, although our investigations in the Abelian case indicate that
at some order in $\al$, modifications would be needed for NNMHV
amplitudes at the least.

\subsubsection*{BCFW}

In section \ref{sec:recursion}, it was shown how in string theory BCFW recursion allows one to write higher point amplitudes in terms of on-shell lower point amplitudes. It was shown quite generally for $4$-point string theory amplitudes how the residue at infinity which could have spoiled recursion is in fact absent. Since the Veneziano
amplitude, for instance, not only diverges but has an
essential singularity, this is perhaps surprising. On the
other hand, intuition says that the behavior at infinity is related
to the UV properties of the theory. Since the string has
(exponentially) good UV properties, recursion is perhaps expected.
From a purely theoretical point of view, the resulting recursion
relations are fascinating as they allow one to reconstruct full amplitudes from their behavior at the boundary of a moduli space. We conjecture that the residue at infinity is absent for any open or closed string theory amplitude at the disk or sphere level in a flat background in any number of space-time dimensions. In the language of \cite{Benincasa:2007xk}, we therefore conjecture that 'tree-level' string theory is constructable.

On a more basic level however, there are several obstructions to be overcome to applying the recursion relations to obtain new amplitudes. In order to calculate in general with the recursion relations one needs to know all three-particle scattering amplitudes for all states in the string theory, and be able to do the spin sums and the sum over the full mass spectrum
\footnote{See e.g. Refs.~\cite{Liu:1987tb,Xiao:2005yn} for results on three- and four-point
superstring amplitudes involving massive string states at the first excited level.}.
Although this problem is tractable in principle, in practice it seems to be very messy. We note for instance that a generic kinematical limit of the $5$-gluon amplitude should, by unitarity, have as a residue a sum over polarizations of $2$ sub-amplitudes. This factorization is however not manifest.

A more modest goal would be to be able to calculate amplitudes order
by order in $\al$. This field theory expansion is not trivial however. This can be seen by the fact that most of the poles at which residues are evaluated move to infinity in this limit, leaving just the massless ones. This needs to be understood much more clearly, especially if one would like to derive the $\al$ corrections to the
MHV amplitude for instance, perhaps by extending the final paragraph of section $5$ to higher point amplitudes. On the other hand, $\al$ corrections to MHV amplitudes are an interesting sandbox to test out ideas as the answer is, with the results of this article, known up to $\al^3$.

Since the residue at infinity vanishes quite naturally for the BCFW two particle shift, it is natural to suspect that more general shifts also have good residue behavior. One example of this is the shift studied in~\cite{Bern:2005hs} for the one loop all-plus gluon amplitude in pure Yang-Mills theory. There a shift was found which displays good residue behavior for the known all-multiplicity expression for the amplitude. Since the $\al^2$ correction to the MHV amplitude is equivalent to this expression save a trivial `$\braket{ij}^4$' prefactor, this will also obey the same recursion relation. However, starting from a known expression limits predictive power. It would still be interesting though to see if a similar procedure would work for our $\al^3$ result in section \ref{sec:alphaprime3}.

\subsection[\protect \phantom{MHV, CSW and BCFW: }field theory structures in string theory amplitudes]{Field theory structures in string theory amplitudes}

We conclude with some more general comments and speculations about field theory structures in string theory amplitudes.

One other shift which could be interesting is the shift in Risager's derivation of the CSW rules~\cite{Risager:2005vk}. Based on the experience with these rules in the Abelian case discussed in this paper in section \ref{sec:derDBI}, it would appear that these are not as straightforward as in ordinary field theory. However, there may very well be some sense to be made of them. It should be noted though that in field theory it is not straightforward to derive the vanishing of the appropriate residues directly~\cite{Elvang:2008na}, despite the fact that this is known to be true from an independent action-based derivation of the $\mathcal{N}=4$ CSW rules~\cite{Boels:2006ir, Boels:2007qn} for all types of external particles.

As is well known, open string amplitudes are directly related to
closed string amplitudes through the KLT relations
\cite{Kawai:1985xq}. Several results in this article therefore
should have a direct interpretation in the closed string. For
instance, one should be able to derive a form of $\al$ corrections
of the MHV gravity amplitudes using these results.

The $\mathcal{N}=2$ string may have some role to play when it comes
to elucidating underlying structures in string theory
amplitudes\footnote{This suggestion was made to us by Shinji
Hirano.}. This was also recently advocated in
\cite{Berkovits:2008ic} with a different motivation. In field
theory, the existence of MHV amplitudes and CSW rules for pure
Yang-Mills theory in four dimensions is related to the existence of
a gauge symmetry on twistor space which is not manifest from the
space-time point of view \cite{Boels:2006ir}. In string theory, the
Ward identity of ordinary gauge symmetry for on-shell amplitudes is
a consequence of the world-sheet conformal symmetry. Extra gauge
symmetry and its accompanying Ward identities should therefore
somehow be connected to a larger conformal algebra, such as that for
the $\mathcal{N}=2$ string. Since ordinary string theory can be
embedded into this theory, this would be interesting to study
further. Supporting evidence for this comes from the fact that the
target space description of ordinary $\mathcal{N}=2$ string theory
is self-dual Yang-Mills theory in $d=4$. As shown in
\cite{Lechtenfeld:2004cc}, there is a direct connection here between
the twistor space gauge symmetry and the $\mathcal{N}=2$ string.

It is interesting to note that the BCFW recursion relations, if
extendable to the full string theory amplitudes as conjectured, basically state
that all amplitudes can be constructed from the three-point
functions and propagators. This is a recurring theme in string
theory, which also can be found for example in open string field
theory, conformal field theory and topological string theory. An
interesting question to ask is whether the connection to these
results goes beyond an analogy. This is of course most natural in
the case of open string field theory, on which we already commented
in the main text. As for connections to conformal field theory and
the topological string, in these theories, as in this case, it is
often easy to split amplitudes into smaller parts, summed over some
intermediate Hilbert space. The hard part is then to identify this
Hilbert space with the space of physical states that one is
interested in. (See e.g.\ section 3.3.2 of \cite{Vonk:2005yv} for
how this is done in the topological string case.) The crucial
ingredient in such proofs lies is the presence of a BRST-like
symmetry on the worldsheet. It would be interesting to see if a
similar structure can be found in the case of the full-fledged open
superstring theory amplitudes. In particular, since this argument is
independent of the global topology of the world-sheet, it would
imply that BCFW-like relations are not only valid at the disk level,
as we argued in this paper, but also at higher loops. It is easy to speculate (and probably hard to prove if true at all) that the BRST-like symmetry mentioned here is connected to the larger gauge symmetry mentioned above for the $\mathcal{N}=2$ string.

As a final comment we would like to point out that most of the basic
ideas used in this paper are actually close to $40$ years old. In a
sense, we simply advocate applying analytic S-matrix techniques to
string theory amplitudes. Motivated by recent developments of
S-matrix methods in field theory, we have added some new
ingredients. These are the spinor helicity notation, the recent
appreciation of the usefulness of complex momenta and a clear focus
on calculating helicity amplitudes. As string theory has until
recently been \emph{the} great success of the original analytic
S-matrix approach, it should perhaps not be too surprising that
these new ingredients appear to be very fruitful to explore. In a
sense, it would appear that string theory is the `simplest quantum field
theory'~\cite{ArkaniHamed:2008gz}.

\acknowledgments The authors gratefully acknowledge discussions with
Paolo di Vecchia, Costas Zoubos, David Skinner and Shinji Hirano.
The research of MV was supported by the South African Research
Chairs Initiative of the Department of Science and Technology and
National Research Foundation. MV would like to thank the Niels Bohr
Institute for kind hospitality during the initial stages of this
project. The work of NO is partially supported by the European
Community`s Human Potential Programme under contract MRTN-CT-2004-005104 `Constituents, fundamental forces and symmetries
of the universe'. The research of RB was supported by a Marie Curie European Reintegration Grant within the 7th European Community Framework Programme. All figures in this paper except figure \ref{fig:schoutenfig} were drawn using Jaxodraw \cite{Binosi:2003yf}.

\appendix

\section{Technical details relevant for section $3$}
%
%
\addtocontents{toc}{\protect\setcounter{tocdepth}{1}}

\subsection{Cyclically reducible polynomials for scalars}
\label{app:proofbasis} In this sub-appendix, we prove that for the
cyclically reducible polynomials of degree $d$ in $N$ scalar
variables, a basis is given by \be
 (a_1 a_2 \ldots a_m) \equiv \sum_{ (i_1, \ldots i_m) } x_{i_1}^{a_1} \cdots x_{i_m}^{a_m} \ ,
\ee where $m \leq N$, $a_i \geq 1$ are positive integers with $\sum
a_i = d$, the sum is over all cyclically ordered sequences $(i_1,
\ldots i_m)$ of length $m$, and we only consider cyclically
inequivalent sequences $(a_1 \ldots a_m)$.

First of all, note that the above polynomials are indeed
cyclically symmetric, and that their reductions after we set several
variables to zero are again cyclically symmetric in the remaining
variables. That is, they are all cyclically reducible. Moreover,
every monomial of degree $d$ appears in one of the polynomials.
Therefore, to show that these polynomials form a basis, we only have
to show that they are the smallest possible ones that are cyclically
reducible.

To see this, take two monomial terms $m_1^{(0)}, m_2^{(0)}$ out
of one of the polynomials above. We now show inductively that
indeed, if one of these terms appears in a cyclically reducible
polynomial, so must the other.
\begin{itemize}
 \item
  In case $m_1^{(0)}$ can be obtained from $m_2^{(0)}$ by a cyclic permutation of the indices, we are done.
 \item
Otherwise, there must be a variable which does not appear in
$m_1^{(0)}$, and one which does not appear in $m_2^{(0)}$. We now
study the monomials $m_1^{(1)}, m_2^{(1)}$ obtained by cyclically
permuting until these `missing variables' have the maximum index
$N$. Clearly, $m_i^{(1)}$ must be in the same basis polynomial as
$m_i^{(0)}$. Now, we take a soft limit on the $N$'th variable in
$m_i^{(1)}$, and go back to step one.
\end{itemize}
As an example, let us study \be
 m_1^{(0)} = x_3^2 x_4 x_5, \qquad m_2^{(0)} = x_1^2 x_3 x_5 \ .
\ee By cyclic permutation in all five variables, the basis
polynomial that contains $m_i^{(0)}$ must also contain $m_i^{(1)}$,
where \be
 m_1^{(1)} = x_1^2 x_2 x_3, \qquad m_2^{(1)} = x_2^2 x_4 x_1 \ .
\ee Note that we have arranged these monomials in such a way that
$x_5$ is missing. When we take the soft limit in $x_5$, we see after
a cyclic permutation in the remaining four variables of $m_2$ that
these basis polynomials must also contain \be
 m_1^{(2)} = x_1^2 x_2 x_3, \qquad m_2^{(2)} = x_3^2 x_1 x_2 \ .
\ee Now we take the soft limit in $x_4$, after which the remaining
monomials are cyclic permutations of each other. Hence, it is clear
that $m_1^{(2)}$ and $m_2^{(2)}$ must be in the same basis
polynomial, and therefore, reasoning backwards, that $m_1^{(0)}$ and
$m_2^{(0)}$ were in the same one. It is easy to see that this
procedure will relate any two monomials that appear in the
polynomials we constructed above, and hence these indeed form a
basis of all cyclically reducible polynomials.

\subsection{Combinatorics of necklaces}
\label{app:necklaces} In section \ref{sec:scalarvariables}, we
encountered the problem of counting the number $N_d$ of necklaces
\be
 (a_1 \ldots a_m) \ ,
\ee with the constraints $a_i \geq 1, \sum a_i = d$. Here, the fact
that the sequence is a `necklace' means that we count sequences
only once if they differ by a cyclic permutation.

The trick to counting these necklaces is to rewrite them as
follows: instead of the number $a_i$, write a one followed by
$(a_i-1)$ zeroes. So, for example, we would rewrite \be
 (3115) \mapsto (1001110000) \ .
\ee A little thought shows that this map is injective onto the set
of `binary' necklaces of length $d$. Furthermore, it is nearly
surjective: the only necklace that is not an image under this map is
the necklace consisting of only zeroes. Therefore, our problem is
solved if we can count binary necklaces of length $d$, where we have
to remember to subtract 1 in the end for the zero necklace.

The latter problem is much easier to solve. Note that there are
$2^d$ binary sequences of length $d$. If a necklace has no symmetry
properties, it appears exactly $d$ times (once for each cyclic
permutation) in the set of binary sequences. So, as a first
approximation, we might take \be
 N_d \sim \frac{2^d}{d} \ .
\ee However, a necklace which consists of a number of copies of a
smaller sequence is undercounted in this way; for example, the
necklace \be
 (1000010000) \ ,
\ee appears only five times in the set of all binary sequences of
length ten, not ten times. Thus, for every divisor $d_i$ of $d$, we
have to compensate for undercounting of the number of repeated
sequences of length $d_i$. Doing this carefully, we find the full
answer, \be
 N_d = -1 + \frac{1}{d} \sum_{d_i | d} \phi(d_i) 2^{d/d_i} \ .
\ee Here, $\phi(d_i)$ is the number of integers $k \leq d_i$ that
are relatively prime to $d_i$. This number is known as Euler's
totient function, and can be written as \be
 \phi(d_i) = d_i \prod_{p | d_i} (1-p^{-1}) \ ,
\ee where the product is over all different primes $p$ dividing
$d_i$.

\subsection{Collinearity of $Q_3^{(N)}$}
\label{app:collinearity} In this sub-appendix, we show that the
expression in the curly brackets of equation
(\ref{eq:collinearlimit}) vanishes after using momentum conservation
and Schouten identities, thus proving that $Q_3^{(N)}$ in equation
(\ref{eq:degsixanswer}) reduces to $Q_3^{(N-1)}$ in the collinear
limit.

\subsubsection*{Terms with metric contractions only}
Let us begin with the terms involving only metric contractions.
Up to a scale factor, these read \bea
 T_1 & = & 3(s_{13}s_{2x}s_{4x}) + 3(s_{12}s_{1x}s_{3x}) + 6(s_{23}s_{1x}s_{4x}) \ret
 && + 3(s_{23}s_{1x}s_{3x}) + 3(s_{24}s_{1x}s_{3x}) + (s_{1x}s_{1x}s_{1x}) \ ,
 \label{eq:T11}
\eea where we use the notation that was introduced below
(\ref{eq:collinearterm}). In order to show that this vanishes, let
us begin by writing down the momentum conservation relation \bea
 0 = \sum_i (s_{i2} s_{1x} s_{3x}) & = & (s_{13} s_{2x} s_{4x}) + (s_{12} s_{1x} s_{3x}) + 2 (s_{23} s_{1x} s_{4x}) \ret
 && + (s_{23} s_{1x} s_{3x}) + (s_{24} s_{1x} s_{3x}) + (s_{2x} s_{1x}
 s_{3x}) \ .
\eea The consecutive terms in this expression arise from the terms
where the index $i$ is smaller than the index represented by $1$,
equal to it, between those represented by 1 and 2, and so on.
Subtracting this equation three times from (\ref{eq:T11}) we see
that the expression for $T_1$ simplifies to \be
 T_1 = (s_{1x}s_{1x}s_{1x}) - 3 (s_{1x} s_{2x} s_{3x}) \ .
 \label{eq:T12}
\ee Next, we use the fact that \be
 \sum_i s_{ix} = (s_{1x}) = 0 \ .
 \label{eq:sums}
\ee Taking the cube of this expression, we find \be
 0 = (s_{1x})^3 = (s_{1x}s_{1x}s_{1x}) + 3 (s_{1x}s_{1x}s_{2x}) + 3(s_{1x} s_{2x} s_{2x}) + 6 (s_{1x}s_{2x}s_{3x}) \ .
 \label{eq:momcubed}
\ee On the other hand, multiplying (\ref{eq:sums}) with $(s_{1x}
s_{2x})$ gives \be
 0 = (s_{1x}) (s_{1x} s_{2x}) =  (s_{1x}s_{1x}s_{2x}) + (s_{1x} s_{2x} s_{2x}) + 3 (s_{1x}s_{2x}s_{3x}) \ .
\ee Subtracting this last equation three times from equation
(\ref{eq:momcubed}), we find back the expression (\ref{eq:T12}), so
we have shown that \be
 T_1 = 0 \ ,
\ee completing the proof for the terms with metric contractions
only.

\subsubsection*{Terms with $\eps$- and metric contractions}
Next, we study the terms which involve both $\eps$- and metric
contractions. These are proportional to \be
 T_2 = (\eps_{x123}s_{2x}) -2(\eps_{234x}s_{1x}) +2(\eps_{123x}s_{3x}) +2(\eps_{134x}s_{2x}) +4(\eps_{123x}s_{4x}) \ .
\ee We begin by comparing this to the following Schouten identity,
where we antisymmetrize over the first five indices: \be
 (\eps_{123x} s_{4x}) - (\eps_{234x} s_{1x}) + (\eps_{134x} s_{2x}) - (\eps_{124x} s_{3x}) = 0 \ .
\ee Subtracting this twice from $T_2$, we are left with \be
 T_2 = (\eps_{x123}s_{2x}) +2(\eps_{123x}s_{3x})  + 2 (\eps_{123x}s_{4x}) + 2 (\eps_{124x} s_{3x}) \ .
\ee Now, we use the momentum conservation result \be
 0 = \sum_i (\eps_{i12x} s_{3x}) = (\eps_{123x} s_{4x}) + (\eps_{123x} s_{3x}) + (\eps_{124x} s_{3x}) \ ,
\ee to reduce this to \be
 T_2 = (\eps_{x123}s_{2x}) \ .
\ee Finally, we see that this remaining term vanishes by noticing
that it equals \be
 0 = \sum_i (\eps_{i12x} s_{2x}) = (\eps_{123x} s_{2x}) \ .
\ee Thus, we have found that also \be
 T_2 = 0 \ ,
\ee and hence the entire polynomial in curly brackets in
(\ref{eq:collinearlimit}) vanishes.

\subsection{Counting Lorentz invariant polynomials}
\label{app:countinv} In this sub-appendix we count the number of
Lorentz-invariants modulo Schouten identities in four dimensions
which can be constructed out of $d$ vectors $p_i$. To avoid
complicated expressions involving the $4$-dimensional epsilon tensor,
we will work in spinor notation. In this notation both epsilon
tensor and metric can be expressed in terms of $\epsilon_{\alpha
\beta}$ and the counting problem reduces to the number of
independent tensors $I(\epsilon)_{\alpha_1 \ldots \alpha_d,
\dalpha_1 \ldots \dalpha_d}$. Hence we can split up the problem into
dotted and undotted indices. Since these obviously transform as a
tensor product of the spinor representation, one way to state the
counting problem is to count the number of independent invariants
(i.e. the $(1)$ representation of $SU(2)$) in the representation
\begin{equation}\label{eq:tensorprodrep}
\underbrace{(2)\otimes (2) \otimes \ldots \otimes (2)}_{d} \ .
\end{equation}
In particular, we find
\begin{eqnarray}
d & & \textrm{reps}\nonumber \\
2 & & (1) \oplus (3) \nonumber \\
4 & & 2 (1) \oplus 2 (3) \oplus (5) \nonumber\\
6 & & 5 (1) \oplus 9 (3) \oplus 5 (5) \oplus (7)\nonumber\\
8 & & 14 (1) \oplus 28 (3) \oplus 20 (5) \oplus 7 (7) \oplus (8) \ .
\end{eqnarray}
Based on these numerics, Sloane's wonderful encyclopedia
\cite{Sloane:seq} suggests Catalan's numbers (A000108) for the
number of trivial representation in the tensor product
\eqref{eq:tensorprodrep}. This can be seen to be true by using one
of the there suggested interpretations of these numbers.
\begin{figure}[t]
  \begin{center}
  \includegraphics[scale=0.4]{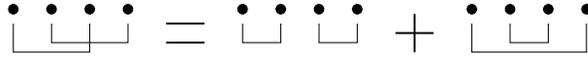}
  \caption{Illustration of the Schouten identity in 2 dimensions. Dots represent indices, brackets represent contraction of the indices by the epsilon tensor.}
  \label{fig:schoutenfig}
  \end{center}
\end{figure}
For this, one should realize Schouten identities are incorporated in
the above counting. For $d=4$ for instance, naive counting yields 3
invariants, one of which is the sum of the others by the Schouten
identity in two dimensions. More generally, this can be illustrated
by considering the construction of the actual invariants. Since
pairs of indices will be contracted for this, one can denote a
particular contraction by drawing lines from one to the other
position. In this notation, it is easy to see that the Schouten
identity can be drawn schematically as in figure
\eqref{fig:schoutenfig} for any four indices. Hence the counting
problem can be reduced to the problem of connecting $d$ dots by
\emph{non-intersecting} pair-wise arcs. The number of these is given
by the Catalan number $c_{\frac{d}{2}}$\footnote{for some visuals,
see for instance
http://www.maths.usyd.edu.au/u/kooc/catalan/cat2smi.pdf}. Having
counted the number of spinor contractions for one kind of spinor, the number of independent
Lorentz invariants follows by squaring.

\section{On-shell recursion relations for the Dirac-Born-Infeld action}\label{app:onshellrecDBI}
In section \ref{sec:action} we managed to derive a significantly simpler
version of perturbation theory for tree level helicity amplitudes
derived from the Dirac-Born-Infeld action in four dimensions.
Furthermore, in section  \ref{sec:recursion} we studied on-shell
recursion relations. It is therefore natural to ask if the DBI
perturbation theory can be used to study some form of on-shell
recursion relations for these amplitudes. This will be discussed in
this appendix.

The simplest example is the 4-point function. It is quite obvious
that shifting in any way will immediately lead to residues at
infinity. Although this can be cured, one could first assume
knowledge of the $4$-point function and see if the 6-point function
can be calculated from this. Therefore, let us study the $6$-point
NMHV amplitude \eqref{eq:6pointDBIampl}. It is natural to try to
shift a positive and a negative helicity photon, since one can
choose then a shift which lets the scalar-to-field strengths
couplings invariant. Let us choose these to be the $3$rd and $4$th
particle respectively and study the amplitude as a contour integral
as in section \ref{sec:recursion},
\begin{equation}
A_6^\textrm{DBI}(1^+ 2^+ 3^+ 4^- 5^- 6^-)(0) = \oint_{z=0}
\frac{A_6^\textrm{DBI}(\ldots \hat{3}^+ \hat{4}^- \ldots)(z)}{z} dz \ ,
\end{equation}
where the hatted indices denote the shifted momenta. Again, we
choose the shift for which $\hat{3}_{\dalpha} = 3_{\dalpha}$ and
$\hat{4}_{\alpha} = 4_{\alpha}$. Pulling the contour to infinity
yields a would-be BCFW recursion relation. However, one can already
see by studying the explicit form of \eqref{eq:6pointDBIampl} that
there are poles in the amplitude where both shifted momenta appear
in the denominator, hence leading to a $z$-independent part of the
amplitude which in turn leads to poles at infinity. Postponing this
point to slightly later, let us calculate one of the finite terms
where for instance particles $3,5,6$ appear on one side of the
recursion,
\begin{equation}\label{eq:onetermin6pointrec}
\textrm{Res} \left(\frac{A_6(z)}{z}, z= \frac{1}{[4|(1 + 2)|3\rangle}
\left(p_1 + p_2 + p_3\right)^2 \right) = \frac{\pi^4 \al^4}{4}
\frac{\braket{1 2}^2 \sbraket{4\hat{\imath}}^2  \braket{\hat{\imath}
3}^2 \sbraket{5 6}^2  }{(p_1 + p_2 + p_4)^2} \ .
 \end{equation}
Here the intermediate, on-shell momentum is
\begin{equation}
\hat{\imath}_{\alpha} \hat{\imath}_{\dot{\alpha}} = {p_1 + p_2 + \hat{p_4}}  \ .
\end{equation}
At this point we can use the standard relation,
\begin{equation}
\sbraket{4 \hat{\imath}}\braket{\hat{\imath} 3} = [4|i|3\rangle  \ ,
\end{equation}
to see that this particular term has the right form to be one of the terms of the expression of the full amplitude~\eqref{eq:6pointDBIampl} .

\begin{figure}[t]
  \begin{center}
  \includegraphics[width=0.9\textwidth]{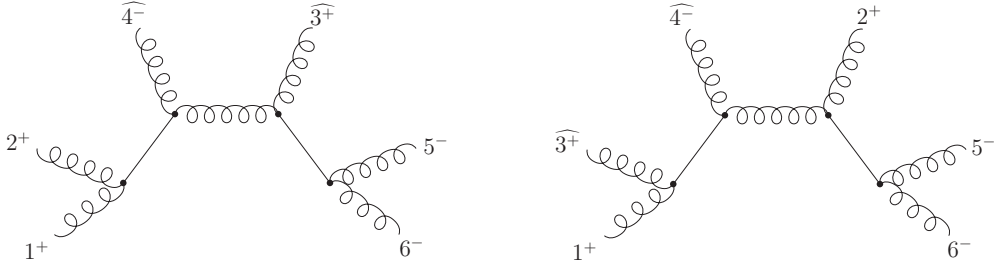}
  \caption{Examples of `good' (left) and `bad' (right) terms in the direct, naive application of BCFW recursion relations for Abelian superstring amplitudes.}
  \label{fig:goodbadtermsrecursion}
  \end{center}
\end{figure}

To see where the missing terms could arise, note that the `good' and
`bad' terms can be drawn by the two example contributions in figure
\ref{fig:goodbadtermsrecursion}: the left figure shows a diagram
which has an explicit kinematic pole from the photon exchange, the
other diagram does not since the scalar propagator is trivial.  If
this propagator would be non-trivial, i.e. contained a standard
momentum factor, then the troublesome terms would also have a pole
at finite values of $z$ (where massive particles go on-shell) and
contribute. Actually, one can promote the scalar to a propagating
field and then expand the resulting expression in $\al$, keeping
only the lowest order term. The trick to circumvent the problem of
the `bad' amplitudes is to first derive the recursion relations and
then expand in $\al$. This is of course in line with general remarks
about the connection between residues at infinity and the UV
behavior of the theory under study in section \ref{sec:recursion}.

Residues of the shifted amplitudes involving the poles of the
now-propagating scalar fields simply reduce to the right expressions
for our particular choice of shift, including kinematic poles since
the rest of the BCFW procedure is then simply a self-similarity
transformation. For instance, to obtain the contribution to the
scattering amplitude with a pole in the channel $(256)$ depicted in
figure \ref{fig:goodbadtermsrecursion} one studies the pole of the
scattering amplitude for which
\begin{equation}
(\hat{p}_3 + p_1)^2 \rightarrow - 1/\al \ ,
\end{equation}
and hence
\begin{equation}
z =  \frac{1}{[4|1|3\rangle} \left(\frac{1}{\al } + (p_3 + p_1 \right)^2) \ .
\end{equation}
Note that this pole migrates to infinity if $\al \rightarrow 0$.
With this simple observation we obtain for this specific
contribution
\begin{equation}
\textrm{Res} \left(\frac{A(z)}{z}, z= \frac{1}{[4|1|3\rangle}
\left(\frac{1}{\al } + (p_3 + p_1)^2\right) \right) = \frac{1}{(p_1
+ p_3)^2 + \frac{1}{\al}} A(\hat{3}^+ 1^+ \hat{k} ) A(\hat{\bar{k}}
\hat{4}^- 2^+ 5^- 6^- ) \ ,
\end{equation}
where $k$ is the now-propagating scalar field with momentum $\hat{p}_3 + p_1$.
As observed before, the hats can simply be removed directly for the
shift under study, while the prefactor can conveniently be expanded
in terms of $\al$. This is easily seen to generate one of the
missing contributions of the six-point amplitude in
\eqref{eq:6pointDBIampl}. Actually, it is diagrammatically
equivalent to one of the Feynman diagrams generated by
\eqref{eq:effectiveDBIaction4D}. It is very important though to
remember that only the leading part in $\al$ of the above
construction actually reproduces the string theory amplitude.

With these extra propagators, it is straightforward to see that
since both a negative and a positive helicity gluon are shifted,
these shifts will never be on the same vertex leading into the
diagram. Therefore, in any expression for the amplitudes derived
from the Feynman rules there are always suppressive powers of $z$,
while for this shift there are no compensating powers of $z$ in the
numerator. Therefore, the residue at infinity of this integral will
always vanish, proving the validity of BCFW recursion formulas in
this case.

\section{Poles and residues for the five-point gluon amplitude}
\label{app:5pointshift}

In this appendix we provide further details of applying the BCFW shift
\eqref{eq:bcfw5gl} to the 5-gluon MHV amplitude \eqref{eq:5gluon}.
First, one finds $s_{23} \longrightarrow s_{23} - z \langle 13 \rangle
[32]$, $s_{51} \longrightarrow s_{51} + z \langle 51 \rangle [25]$
and $\langle 23 \rangle \longrightarrow \langle 23 \rangle - z
\langle 13 \rangle$. Thus, the analytically continued amplitude is
{\footnotesize
\begin{equation} A_5(z) = \frac{\langle 12
\rangle^4}{\langle 12 \rangle \langle 23 \rangle \langle 34 \rangle
\langle 45 \rangle \langle 51 \rangle} \big( \alpha'^2 (s_{51} +
z\langle 51 \rangle [25]) s_{23} f_1(z) + \alpha'^2 [12] \langle 23
\rangle [35] \langle 51 \rangle f_2(z) \big) \ , \label{eq:shiftedA5}
\end{equation}}
where
{\footnotesize \begin{eqnarray} f_1(z) &=&
\frac{\Gamma(\alpha' s_{23} - \alpha'z\langle 13 \rangle [32])
\Gamma(\alpha' s_{51} + \alpha' z \langle 51 \rangle [25])
\Gamma(\alpha' s_{34} + 1) \Gamma(\alpha' s_{45}+1)}{\Gamma(\alpha'
s_{23} + \alpha' s_{34} - \alpha' z \langle 13 \rangle [32] + 1)
\Gamma(\alpha' s_{51} + \alpha'
s_{45} + \alpha' z \langle 51 \rangle [25] + 1)} \nonumber \\
&\phantom{=}& \hspace{0.5cm} \times \phantom{.}_3 F_2
\left[ \begin{array}{c} \alpha' s_{23} - \alpha' z\langle 13 \rangle [32], \alpha' s_{51} + \alpha' z\langle 51 \rangle [25], -\alpha' s_{35} \\
\alpha' s_{23} + \alpha' s_{34} - \alpha' z\langle 13 \rangle [32] +
1, \alpha' s_{51} + \alpha' s_{45} + \alpha' z\langle 51 \rangle
[25] + 1 \end{array} ; 1\right] \ ,
\end{eqnarray}}
and {\footnotesize \begin{eqnarray} f_2(z) &=& \frac{\Gamma(\alpha'
s_{23}- \alpha' z\langle 13 \rangle [32]+1) \Gamma(\alpha' s_{51}+
\alpha' z\langle 51 \rangle [25]+1) \Gamma(\alpha' s_{34} + 1)
\Gamma(\alpha' s_{45}+1)}{\Gamma(\alpha' s_{23} + \alpha' s_{34} -
\alpha' z\langle 13 \rangle [32] + 2) \Gamma(\alpha' s_{51} +
\alpha'
s_{45} + \alpha' z\langle 51 \rangle [25] + 2)} \nonumber \\
&\phantom{=}& \hspace{0.5cm} \times \phantom{.}_3 F_2 \left[
\begin{array}{c} \alpha' s_{23} - \alpha' z\langle 13 \rangle [32] + 1, \alpha' s_{51} + \alpha' z\langle 51 \rangle [25] +1, -\alpha'
s_{35}+1 \\ \alpha' s_{23} + \alpha' s_{34} - \alpha' z\langle 13
\rangle [32] + 2, \alpha' s_{51} + \alpha' s_{45} + \alpha' z\langle
51 \rangle [25] + 2 \end{array} ; 1\right] \ .
\end{eqnarray}}
As a function of $z$, $\frac{A_5(z)}{z}$ has the following poles and
corresponding residues:
\begin{itemize}
\item at $z=\frac{\langle 23 \rangle}{\langle 13 \rangle}$, with
residue {\footnotesize \begin{eqnarray}
&\phantom{=}& \hspace{-1.7cm} \mathrm{Res}  \left(\textstyle{\frac{A_5(z)}{z}}, z=\textstyle{\frac{\langle 23 \rangle}{\langle 13 \rangle}} \right) \nonumber \\
&=& -\frac{\langle 12 \rangle^4}{\langle 12 \rangle \langle 23
\rangle \langle 34 \rangle \langle 45 \rangle \langle 51 \rangle}
\frac{\Gamma \left( \alpha' s_{51} + \alpha'
\textstyle{\frac{\langle 23 \rangle \langle 51 \rangle [25]}{\langle
13 \rangle}}  + 1 \right) \Gamma(\alpha' s_{45} + 1)}{ \Gamma \left(
\alpha' s_{51} + \alpha' s_{45} + \alpha' \textstyle{\frac{\langle
23 \rangle \langle 51 \rangle
[25]}{\langle 13 \rangle}} + 1 \right)} \label{eq:A5factorizes}\\
&=& -\frac{\langle 12 \rangle^4}{\langle 12 \rangle \langle 23
\rangle \langle 34 \rangle \langle 45 \rangle \langle 51 \rangle}
\left[ 1 - \textstyle{\frac{\pi^2 \alpha'^2}{6}} s_{45} \left(
s_{51} + \textstyle{\frac{\langle 23 \rangle \langle 51 \rangle
[25]}{\langle 13 \rangle}} \right) + \mathcal{O}(\alpha'^3) \right] \ .
\label{eq:residue1}
\end{eqnarray}}
As opposed to the 4-gluon case \eqref{eq:A4-gluonexchange} this does
not reproduce the Yang-Mills result. Of course, this follows simply
from the fact that on this pole the amplitude factorizes into a
3-gluon amplitude and a 4-gluon amplitude (as can be seen directly
in \eqref{eq:A5factorizes}), where the latter amplitude receives
$\alpha'$-corrections in superstring theory.

\item at $z = \textstyle{\frac{\alpha' s_{23} + k_1}{\alpha'
\langle 13 \rangle [32]}}$, $k_1 \in \mathbb{N}$, with residue
{\footnotesize
\begin{eqnarray} &\phantom{=}& \hspace{-1.7cm} \mathrm{Res}
\left(\textstyle{\frac{A_5(z)}{z}}, z=\textstyle{
\frac{\alpha' s_{23} + k_1}{\alpha' \langle 13 \rangle [32]}} \right) \nonumber \\
&=& \frac{\langle 12 \rangle^4}{\langle 12 \rangle \langle 23
\rangle \langle 34 \rangle \langle 45 \rangle \langle 51 \rangle}
\left[ \alpha'^2 \left( s_{51} +
\left(s_{23}+\textstyle{\frac{k_1}{\alpha'}} \right)
\textstyle{\frac{\langle 51 \rangle [25]}{\langle 13 \rangle [32]}}
\right) s_{23} \widetilde{f_1} + \alpha'^2 [12] \langle 23 \rangle
[35] \langle 51 \rangle \widetilde{f_2} \right] \ ,
\label{eq:fullresidueA5(1)}
\end{eqnarray}}
where {\footnotesize \begin{eqnarray} \widetilde{f_1} &=&
\frac{(-1)^{k_1+1}}{k_1!} \frac{1}{\alpha' s_{23} + k_1}
\frac{\Gamma\left( \alpha' s_{51} + (\alpha' s_{23} +
k_1)\textstyle{\frac{\langle 51 \rangle [25]}{\langle 13 \rangle
[32]}} \right) \Gamma(\alpha' s_{34} + 1)\Gamma(\alpha'
s_{45}+1)}{\Gamma(-k_1 + 1 + \alpha' s_{34}) \Gamma\left(\alpha'
s_{51} + \alpha' s_{45} + (\alpha' s_{23}+
k_1)\textstyle{\frac{\langle 51 \rangle
[25]}{\langle 13 \rangle [32]}} + 1\right)} \nonumber \\
&\phantom{=}& \times \phantom{.}_3 F_2 \left[
\begin{array}{c} -k_1, \alpha' s_{51} + (\alpha' s_{23}+
k_1)\textstyle{\frac{\langle 51 \rangle [25]}{\langle 13 \rangle
[32]}}, -\alpha' s_{35}
\\ -k_1 + 1 + \alpha' s_{34}, \alpha' s_{51} + \alpha' s_{45} + \big(\alpha' s_{23}+
k_1 \big)\textstyle{\frac{\langle 51 \rangle [25]}{\langle 13
\rangle [32]}} + 1
\end{array} ; 1\right] \\
&=& \frac{1}{k_1^3} \left( \frac{s_{34} \langle 13 \rangle
[32]}{\langle 51 \rangle [25]} + \frac{s_{35}}{1 + \frac{\langle 51
\rangle [25]}{\langle 13 \rangle [32]}} \right)\alpha' +
\mathcal{O}(\alpha'^2) \ ,
\end{eqnarray}}
and {\footnotesize \begin{eqnarray} \widetilde{f_2} &=&
\frac{(-1)^{k_1}}{(k_1-1)!} \frac{1}{\alpha' s_{23} + k_1}
\frac{\Gamma\left( \alpha' s_{51} + (\alpha' s_{23} +
k_1)\textstyle{\frac{\langle 51 \rangle [25]}{\langle 13 \rangle
[32]}} +1 \right) \Gamma(\alpha' s_{34} + 1)\Gamma(\alpha'
s_{45}+1)}{\Gamma(-k_1 + 2 + \alpha' s_{34}) \Gamma\left(\alpha'
s_{51} + \alpha' s_{45} + (\alpha' s_{23}+
k_1)\textstyle{\frac{\langle 51 \rangle
[25]}{\langle 13 \rangle [32]}} + 2\right)} \nonumber \\
&\phantom{=}& \times \phantom{.}_3 F_2 \left[
\begin{array}{c} -k_1+1, \alpha' s_{51} + (\alpha' s_{23}+
k_1)\textstyle{\frac{\langle 51 \rangle [25]}{\langle 13 \rangle
[32]}}+1, -\alpha' s_{35}+1 \\ -k_1 + 2 + \alpha' s_{34}, \alpha'
s_{51} + \alpha' s_{45} + \big(\alpha' s_{23}+ k_1
\big)\textstyle{\frac{\langle 51 \rangle [25]}{\langle 13 \rangle
[32]}} + 2
\end{array} ; 1\right] \\
&=& - \frac{1}{k_1^2} \frac{1}{1 + \frac{\langle 51 \rangle
[25]}{\langle 13 \rangle [32]}} + \mathcal{O}(\alpha') \ ,
\end{eqnarray}}
leading to the following $\alpha'$ expansion of the residue at $z =
\textstyle{\frac{\alpha' s_{23} + k_1}{\alpha' \langle 13 \rangle
[32]}}$ {\scriptsize \begin{eqnarray} &\phantom{=}& \hspace{-1cm}
\mathrm{Res} \left(\textstyle{\frac{A_5(z)}{z}}, z=\textstyle{
\frac{\alpha' s_{23} + k_1}{\alpha' \langle 13 \rangle [32]}}
\right) \nonumber = \frac{\alpha'^2}{k_1^2} \frac{\langle 12
\rangle^4}{\langle 12 \rangle \langle 23 \rangle \langle 34 \rangle
\langle 45 \rangle \langle 51 \rangle} \left[ s_{23} \frac{\langle
51 \rangle [25]}{\langle 13 \rangle [32]} \left( \frac{s_{34}
\langle 13 \rangle [32]}{\langle 51 \rangle [25]} + \frac{s_{35}}{1
+ \frac{\langle 51 \rangle [25]}{\langle 13 \rangle [32]}} \right)
\right.
\nonumber \\
&\phantom{=}& \hspace{6.6cm} \left. - [12]\langle 23 \rangle [35]
\langle 51 \rangle \frac{1}{1 + \frac{\langle 51 \rangle
[25]}{\langle 13 \rangle [32]}} \right] + \mathcal{O}(\alpha'^3) \ .
\label{eq:residue2}
\end{eqnarray}}

\item at $z = \textstyle{\frac{\alpha' s_{51} + k_2}{\alpha'
\langle 15 \rangle [25]}}$, $k_2 \in \mathbb{N}$, with residue
{\footnotesize
\begin{equation} \mathrm{Res}
\left(\textstyle{\frac{A_5(z)}{z}}, z=\textstyle{ \frac{\alpha'
s_{51} + k_2}{\alpha' \langle 15 \rangle [25]}} \right) =
\frac{\langle 12 \rangle^4}{\langle 12 \rangle \langle 23 \rangle
\langle 34 \rangle \langle 45 \rangle \langle 51 \rangle} \left[
-\alpha' k_2 s_{23} \widetilde{f_1} + \alpha'^2 [12] \langle 23
\rangle [35] \langle 51 \rangle \widetilde{f_2} \right] \ ,
\label{eq:fullresidueA5(2)}
\end{equation}}
where {\footnotesize \begin{eqnarray} \widetilde{f_1} &=&
\frac{(-1)^{k_2+1}}{k_2!} \frac{1}{\alpha' s_{51} + k_2}
\frac{\Gamma\left( \alpha' s_{23} - (\alpha' s_{51} +
k_2)\textstyle{\frac{\langle 13 \rangle [32]}{\langle 15 \rangle
[25]}} \right) \Gamma(\alpha' s_{34} + 1)\Gamma(\alpha'
s_{45}+1)}{\Gamma(-k_2 + 1 + \alpha' s_{45}) \Gamma\left(\alpha'
s_{23} + \alpha' s_{34} - (\alpha' s_{51}+
k_2)\textstyle{\frac{\langle 13 \rangle
[32]}{\langle 15 \rangle [25]}} + 1\right)} \nonumber \\
&\phantom{=}& \times \phantom{.}_3 F_2 \left[
\begin{array}{c} \alpha' s_{23} - (\alpha' s_{51}+
k_2)\textstyle{\frac{\langle 13 \rangle [32]}{\langle 15 \rangle
[25]}},-k_2, -\alpha' s_{35}
\\ \alpha' s_{23} + \alpha' s_{34} - \big(\alpha' s_{51}+
k_2 \big)\textstyle{\frac{\langle 13 \rangle [32]}{\langle 15
\rangle [25]}} + 1, -k_2 + 1 + \alpha' s_{45}
\end{array} ; 1\right] \\
&=& \frac{1}{k_2^3} \left( -\frac{s_{45} \langle 15 \rangle
[25]}{\langle 13 \rangle [32]} + \frac{s_{35}}{1 - \frac{\langle 13
\rangle [32]}{\langle 15 \rangle [25]}} \right)\alpha' +
\mathcal{O}(\alpha'^2) \ ,
\end{eqnarray}}
and {\footnotesize \begin{eqnarray} \widetilde{f_2} &=&
\frac{(-1)^{k_2}}{(k_2-1)!} \frac{1}{\alpha' s_{51} + k_2}
\frac{\Gamma\left( \alpha' s_{23} - (\alpha' s_{51} +
k_2)\textstyle{\frac{\langle 13 \rangle [32]}{\langle 15 \rangle
[25]}} +1 \right) \Gamma(\alpha' s_{34} + 1)\Gamma(\alpha'
s_{45}+1)}{\Gamma(-k_2 + 2 + \alpha' s_{45}) \Gamma\left(\alpha'
s_{23} + \alpha' s_{34} - (\alpha' s_{51}+
k_2)\textstyle{\frac{\langle 13 \rangle
[32]}{\langle 15 \rangle [25]}} + 2\right)} \nonumber \\
&\phantom{=}& \times \phantom{.}_3 F_2 \left[
\begin{array}{c} \alpha' s_{23} - (\alpha' s_{51}+
k_2)\textstyle{\frac{\langle 13 \rangle [32]}{\langle 15 \rangle
[25]}}+1,-k_2+1, -\alpha' s_{35}+1
\\ \alpha' s_{23} + \alpha' s_{34} - \big(\alpha' s_{51}+
k_2 \big)\textstyle{\frac{\langle 13 \rangle [32]}{\langle 15
\rangle [25]}} + 2, -k_2 + 2 + \alpha' s_{45}
\end{array} ; 1\right] \\
&=& - \frac{1}{k_2^2} \frac{1}{1 - \frac{\langle 13 \rangle
[32]}{\langle 15 \rangle [25]}} + \mathcal{O}(\alpha') \ ,
\end{eqnarray}}
leading to the following $\alpha'$ expansion of the residue at $z =
\textstyle{\frac{\alpha' s_{51} + k_2}{\alpha' \langle 15 \rangle
[25]}}$ {\scriptsize \begin{eqnarray} &\phantom{=}& \hspace{-1cm}
\mathrm{Res} \left(\textstyle{\frac{A_5(z)}{z}}, z=\textstyle{
\frac{\alpha' s_{51} + k_2}{\alpha' \langle 15 \rangle [25]}}
\right) \nonumber = \frac{\alpha'^2}{k_2^2} \frac{\langle 12
\rangle^4}{\langle 12 \rangle \langle 23 \rangle \langle 34 \rangle
\langle 45 \rangle \langle 51 \rangle} \left[ s_{23} \left(
\frac{s_{45} \langle 15 \rangle [25]}{\langle 13 \rangle
[32]} - \frac{s_{35}}{1- \frac{\langle 13 \rangle [32]}{\langle 15 \rangle [25]}} \right) \right. \nonumber \\
&\phantom{=}& \hspace{6.6cm} \left. - [12]\langle 23 \rangle [35]
\langle 51 \rangle \frac{1}{1 - \frac{\langle 13 \rangle
[32]}{\langle 15 \rangle [25]}} \right] + \mathcal{O}(\alpha'^3) \ .
\label{eq:residue3}
\end{eqnarray}}
\end{itemize}
Adding the residues (\ref{eq:residue1}), (\ref{eq:residue2}),
(\ref{eq:residue3}) one obtains\footnote{Using the identity
$[12]\langle 23 \rangle [35] \langle 51 \rangle = \frac{1}{2} \big(
s_{23} s_{34} + s_{45} s_{51} - s_{12} s_{23} - s_{34} s_{45} -
s_{51} s_{12} \big) - 2i\epsilon(1,2,3,4)$.} {\scriptsize
\begin{eqnarray} &\phantom{=}& \hspace{-1.2cm} \mathrm{Res}
\left(\textstyle{\frac{A_5(z)}{z}}, z=\textstyle{\frac{\langle 23
\rangle}{\langle 13 \rangle}} \right) + \sum_{k_1=1}^\infty
\mathrm{Res} \left(\textstyle{\frac{A_5(z)}{z}}, z=\textstyle{
\frac{\alpha' s_{23} + k_1}{\alpha' \langle 13 \rangle [32]}}
\right) + \sum_{k_2 = 1}^\infty \mathrm{Res}
\left(\textstyle{\frac{A_5(z)}{z}}, z=\textstyle{ \frac{\alpha'
s_{51} + k_2}{\alpha' \langle 15 \rangle [25]}} \right) \nonumber \\
&\phantom{=}& \hspace{-1.1cm}= -\frac{\langle 12 \rangle^4}{\langle
12 \rangle \langle 23 \rangle \langle 34 \rangle \langle 45 \rangle
\langle 51 \rangle} \left( 1 - \frac{\pi^2 \alpha'^2}{12} \Big(
s_{12} s_{23} + s_{23} s_{34} + s_{34} s_{45} + s_{45} s_{51} +
s_{51} s_{12} + 4i\epsilon(1,2,3,4) \Big) + \mathcal{O}(\alpha'^3)
\right) \ ,
\end{eqnarray}}
\hspace{-2.2mm} which is consistent with \cite{Stieberger:2006te},
eqs. (37) and (39). In particular this implies that we have explicitly
shown that up to order $\mathcal{O}(\alpha'^3)$ the residue at infinity
is absent,
\begin{equation}
\oint_\infty \frac{A_5(z)}{z} dz = \mathcal{O}(\alpha'^3) \ .
\label{eq:no-obstruction-for-A5-RR}
\end{equation}
The argument in the text proves this to all orders.

\bibliographystyle{JHEP}
\bibliography{biblio}

\end{document}